\documentclass[english]{article}
\usepackage[T1]{fontenc}
\usepackage[latin9]{inputenc}
\usepackage{refstyle}
\usepackage{float}
\usepackage{rotfloat}
\usepackage{booktabs}
\usepackage{amsmath}
\usepackage{graphicx}
\usepackage[numbers,super,sort&compress]{natbib}

\makeatletter

\AtBeginDocument{\providecommand\eqref[1]{\ref{eq:#1}}}
\AtBeginDocument{\providecommand\subsecref[1]{\ref{subsec:#1}}}
\AtBeginDocument{\providecommand\tabref[1]{\ref{tab:#1}}}
\AtBeginDocument{\providecommand\secref[1]{\ref{sec:#1}}}
\AtBeginDocument{\providecommand\figref[1]{\ref{fig:#1}}}
\providecommand{\tabularnewline}{\\}
\floatstyle{ruled}
\newfloat{algorithm}{tbp}{loa}
\providecommand{\algorithmname}{Algorithm}
\floatname{algorithm}{\protect\algorithmname}
\RS@ifundefined{subsecref}
  {\newref{subsec}{name = \RSsectxt}}
  {}
\RS@ifundefined{thmref}
  {\def\RSthmtxt{theorem~}\newref{thm}{name = \RSthmtxt}}
  {}
\RS@ifundefined{lemref}
  {\def\RSlemtxt{lemma~}\newref{lem}{name = \RSlemtxt}}
  {}

\newcommand{\lyxaddress}[1]{
	\par {\raggedright #1
	\vspace{1.4em}
	\noindent\par}
}
\newcommand{\lyxrightaddress}[1]{
	\par {\raggedleft \begin{tabular}{l}\ignorespaces
	#1
	\end{tabular}
	\vspace{1.4em}
	\par}
}

\usepackage[version=3]{mhchem}
\usepackage{braket}
\usepackage{url}
\usepackage{notes2bib}

\@ifundefined{showcaptionsetup}{}{%
 \PassOptionsToPackage{caption=false}{subfig}}
\usepackage{subfig}
\makeatother

\usepackage{babel}
\begin{document}
\title{Fully numerical Hartree--Fock and density functional calculations.
II. Diatomic molecules}
\author{Susi Lehtola}
\maketitle

\lyxaddress{Department of Chemistry, University of Helsinki, P.O. Box 55 (A.
I. Virtasen aukio 1), FI-00014 University of Helsinki, Finland}

\lyxrightaddress{susi.lehtola@alumni.helsinki.fi}
\begin{abstract}
We present the implementation of a variational finite element solver
in the \textsc{HelFEM} program for benchmark calculations on diatomic
systems. A basis set of the form $\chi_{nlm}(\mu,\nu,\phi)=B_{n}(\mu)Y_{l}^{m}(\nu,\phi)$
is used, where $(\mu,\nu,\phi)$ are transformed prolate spheroidal
coordinates, $B_{n}(\mu)$ are finite element shape functions, and
$Y_{l}^{m}$ are spherical harmonics. The basis set allows for an
arbitrary level of accuracy in calculations on diatomic molecules,
which can be performed at present with either nonrelativistic Hartree--Fock
(HF) or density functional (DF) theory. Hundreds of DFs at the local
spin-density approximation (LDA), generalized gradient approximation
(GGA) and the meta-GGA level can be used through an interface with
the \textsc{Libxc} library; meta-GGA and hybrid DFs aren't available
in other fully numerical diatomic program packages. Finite electric
fields are also supported in \textsc{HelFEM}, enabling access to electric
properties.

We introduce a powerful tool for adaptively choosing the basis set
by using the core Hamiltonian as a proxy for its completeness. \textsc{HelFEM}
and the novel basis set procedure are demonstrated by reproducing
the restricted open-shell HF limit energies of 68 diatomic molecules
from the $1^{\text{st}}$ to the $4^{\text{th}}$ period with excellent
agreement with literature values, despite requiring \emph{orders of
magnitude} fewer parameters for the wave function. Then, the electric
properties of the \ce{BH} and \ce{N2} molecules under finite field
are studied, again yielding excellent agreement with previous HF limit
values for energies, dipole moments, and dipole polarizabilities,
again with much more compact wave functions than what were needed
in the literature references. Finally, HF, LDA, GGA, and meta-GGA
calculations of the atomization energy of \ce{N2} are performed,
demonstrating the superb accuracy of the present approach.
\end{abstract}
\global\long\def\ERI#1#2{(#1|#2)}%
\global\long\def\bra#1{\Bra{#1}}%
\global\long\def\ket#1{\Ket{#1}}%
\global\long\def\braket#1{\Braket{#1}}%

\newcommand*\citeref[1]{ref. \citenum{#1}} 
\newcommand*\Citeref[1]{Ref. \citenum{#1}} 
\newcommand*\citerefs[1]{refs. \citenum{#1}} 


\section{Introduction\label{sec:Introduction}}

In the first part of this series,\citep{Lehtola2019a} we presented
fully numerical Hartree--Fock (HF) and density functional calculations
on atoms. The present manuscript focuses on diatomic molecules, which
may serve as even more stringent tests of \emph{ab initio} as well
as density functional methods than atoms, as the methods' accuracy
can be probed for multiple kinds of covalent bonds, as well as for
ionic bonds and dispersion effects. As reviewed at length in \citeref{Lehtola2019c},
the history of fully numerical calculations on diatomic molecules
is almost as long as that on atoms, starting with the partial wave
approach of McCullough.\citep{McCullough1974,McCullough1975} The
partial wave approach appears to have since fallen out of use, with
the well-known programs by Heinemann, Laaksonen, Sundholm, Kobus and
coworkers\citep{Heinemann1988,Sundholm1989,Kobus1996,Kobus2013} relying
on grid-based approaches for the angular expansion.

However, it is not clear whether the partial wave approach should
be fully forgotten: on the contrary, there is a strong argument for
resuscitating its use. Unlike in the fully grid-based approaches (see
references in \citeref{Lehtola2019c}), in the partial wave approach
angular integrals can be evaluated analytically in closed form, requiring
no approximations or truncations to be made in the numerical implementation
beyond the choice of the single-particle basis set. This is especially
true for the two-electron integrals: as with the Legendre expansion
in the atomic case,\cite{Lehtola2019a} the angular integrals arising
for diatomic molecules from the Neumann expansion\citep{Rudenberg1951}
can be performed analytically within the partial wave expansion,\citep{McCullough1975}
indicating that such a basis is extremely convenient for calculations.

While a suitable variational re-implementation of McCullough's partial-wave
approach has been reported in \citeref{Artemyev2004} employing a
basis spline approach, the calculations therein were limited to first
and second period atoms and diatomic molecules, leaving it unclear
whether the approach is tractable for heavier systems. As we were
furthermore unable to obtain a copy of the program of \citeref{Artemyev2004},
we decided to write a new finite element program from scratch, employing
modern programming paradigms and libraries. Some omissions in the
equations of \citeref{Artemyev2004} were thereby found, as shall
be described below. We have also developed faster algorithms for the
formation of the Coulomb and exchange matrices, which do not appear
to have been used in \citeref{Artemyev2004}. Finally, unlike the
program of \citeref{Artemyev2004}, the present implementation is
parallellized, and also supports calculations within density functional
theory\citep{Hohenberg1964,Kohn1965} (DFT) in addition to the HF
calculations of \citeref{Artemyev2004}.

In the present work, we will thus describe the implementation of a
finite element solver for HF and DFT on diatomic molecules, employing
the partial wave approach originally proposed by McCullough.\citep{McCullough1974,McCullough1975}
The program is called \textsc{HelFEM}\citep{HelFEM} for the city
and university of Helsinki where the present author is situated, and
for the electronic Hamiltonian
\begin{equation}
\hat{H}_{el}=-\frac{1}{2}\sum_{i}\nabla_{i}^{2}-\sum_{i}\frac{Z_{A}}{r_{iA}}-\sum_{i}\frac{Z_{B}}{r_{iB}}+\sum_{i>j}\frac{1}{r_{ij}}+\frac{Z_{A}Z_{B}}{r_{AB}},\label{eq:diatH}
\end{equation}
and for the finite element method (FEM). \textsc{HelFEM} is open source
(GNU General Public License) and it has been written in object-oriented
C++. \textsc{S}everal recently published open source algorithms and
libraries are used in \textsc{HelFEM}. \textsc{HelFEM} is especially
linked to the \textsc{Libxc} library,\citep{Lehtola2018} which provides
hundreds of local spin density approximation\citep{Kohn1965} (LDA),
generalized-gradient approximation\citep{Langreth1980} (GGA) as well
as meta-GGA\citep{Perdew1999} exchange-correlation functionals. In
contrast, the programs by Heinemann, Laaksonen, Sundholm, Kobus and
coworkers\citep{Heinemann1988,Sundholm1989,Kobus1996,Kobus2013} are
limited to few LDA and GGA functionals; to our knowledge, meta-GGAs
have not been previously available in all-electron diatomic programs.

Both pure and global hybrid density functionals are supported in \textsc{HelFEM};
alike the atomic calculations discussed in \citeref{Lehtola2019a},
range-separated functionals are not yet supported for obvious reasons
that are discussed below. Both spin-restricted, spin-restricted open-shell,
as well as spin-unrestricted calculations are supported in \textsc{HelFEM}.
As far as we know, spin-unrestricted all-electron real-space calculations
on diatomics have only been reported so far by Heinemann and coworkers.\citep{Heinemann1990a,Heinemann1993}

As was discussed in part I for the atomic calculations,\cite{Lehtola2019a}
the data layout in \textsc{HelFEM} is deliberately similar to what
is used in typical Gaussian-basis quantum chemistry programs; this
also holds in the case of diatomic calculations. Thanks to this, many
functionalities, such as the DIIS\citep{Pulay1980,Pulay1982} and
ADIIS\citep{Hu2010} self-consistent field (SCF) procedure convergence
accelerators have been adopted directly from the \textsc{Erkale} program.\citep{erkale,Lehtola2012}
In addition, interfaces to multiconfigurational methods,\citep{Hinze1973}
configuration interaction, and coupled-cluster theories\citep{Cizek1966}
available in \emph{e.g.} \textsc{Psi4}\citep{Parrish2017} or \textsc{PySCF}\citep{Sun2018}
could be implemented in the future.

Unlike the commonly used programs for fully numerical calculations
on diatomic molecules by Heinemann, Laaksonen, Sundholm, Kobus and
coworkers,\citep{Heinemann1988,Sundholm1989,Kobus1996,Kobus2013}
\textsc{HelFEM} calculates the Coulomb and exchange matrices in the
``traditional'' manner with two-electron integrals. This means that
the value for the practical infinity $r_{\infty}$ can be determined
by the behavior of the electron density alone. Furthermore, the approach
in \textsc{HelFEM} is strictly variational: the energies given by
the program are true upper bounds to the energy computed in a complete
basis set (CBS). This can be contrasted to the energies produced \emph{e.g.}
by the \textsc{x2dhf} program,\citep{Kobus1996,Kobus2013} which are
typically antivariational, \emph{i.e.} the energy approaches the converged
value from below due to inaccuracies in the potential. Because, for
instance, the HF energy is an upper bound to the energy of the \emph{many-electron}
wave function, it is clearly beneficial if the HF energy itself is
also estimated variationally by the numerical approach. Although both
numerical approaches give the same solution at convergence,\citep{Kobus1993}
variationality makes it easier to establish convergence to the basis
set limit.

Furthermore, the approach in \textsc{HelFEM} guarantees smooth and
rapid convergence of the SCF procedure without the need to adjust
relaxation parameters as in \textsc{x2dhf}. An initial guess wave
function or orbital symmetries are also unnecessary in \textsc{HelFEM},
unlike for \textsc{x2dhf} or the program by Heinemann and coworkers;
this again greatly simplifies running calculations. Finally, unlike
the finite difference approach used in \textsc{x2dhf} where smaller
grid spacings radically increase the number of steps to solution,
the speed of SCF convergence in \textsc{HelFEM} is not affected by
the size of the basis set. Although the diagonalization cost is affected
by the use of a larger basis set, it is not rate determining in our
calculations.

The layout of the article is the following. Next, in the Theory section,
we will present all the equations that are necessary for a finite
element implementation of the partial-wave approach for diatomic molecules,
as well as present a novel adaptive approach for choosing the basis
set cost-efficiently for diatomic calculations. The Theory section
is followed by a Computational Details section, which describes the
present implementation and details various convergence parameters
that were used for the calculations. The Results section shows three
applications of the novel program: the calculation of restricted open-shell
HF (ROHF) limit ground state energies of 70 diatomic molecules from
\citerefs{Jensen2005} and \citenum{Williams2008} that range from
the $1^{\text{st}}$ to the $4^{\text{th}}$ period, the finite field
electric properties of the \ce{BH} and \ce{N2} molecules at the
HF limit, and the atomization energy of \ce{N2} with HF, as well
as LDA, GGA, and meta-GGA functionals. The article ends with brief
Summary and Conclusions sections.

As the article relies on knowledge on the finite element approach
that was presented in the first part of this series,\cite{Lehtola2019a}
it should be read first. Atomic units are used unless specified otherwise.
The Einstein summation convention is used, meaning summations are
implied over repeated indices.

\section{Theory\label{sec:Theory}}

\subsection{Coordinate system\label{subsec:Coordinate-system}}

Modified prolate spheroidal coordinates $(\mu,\nu,\phi)$ due to Becke\citep{Becke1982}
(see illustrations in \citeref{Lehtola2019c}) are used:
\begin{align}
x= & R_{h}\sinh\mu\sin\nu\cos\phi,\label{eq:x-eq}\\
y= & R_{h}\sinh\mu\sin\nu\sin\phi,\label{eq:y-eq}\\
z= & R_{h}\cosh\mu\cos\nu,\label{eq:z-eq}
\end{align}
where, for convenience, we have defined the half-bond distance 
\begin{equation}
R_{h}=\frac{1}{2}R\label{eq:Rhalf}
\end{equation}
to avoid carrying various fractions in the equations. In \eqref{Rhalf},
$R$ is the internuclear distance, the two nuclei $Z_{A}$ and $Z_{B}$
being thus placed at $(0,0,-R_{h})$ and $(0,0,R_{h})$, respectively.
The distances of a point from the two nuclei and from the origin can
thereby be written as
\begin{align}
r_{A}= & R_{h}(\cosh\mu+\cos\nu),\label{eq:r1}\\
r_{B}= & R_{h}(\cosh\mu-\cos\nu),\label{eq:r2}\\
r= & R_{h}\sqrt{\cosh^{2}\mu+\cos^{2}\nu-1},\label{eq:r}
\end{align}
respectively, while the spherical polar angle $\cos\theta=z/r$ can
be written as

\begin{equation}
\cos\theta=\frac{\cosh\mu\cos\nu}{\sqrt{\cosh^{2}\mu+\cos^{2}\nu-1}}.\label{eq:costh}
\end{equation}
As was discussed in \citeref{Lehtola2019c}, isosurfaces of $\mu$
approach spheres for large values of the coordinate. This can also
be seen from \eqref{r}: for large values of $\mu$, the distance
from the origin approaches
\begin{equation}
r\to R_{h}\cosh\mu,\label{eq:r-limit}
\end{equation}
the same limit is also achieved when $R\to0$, in which case the usual
spherical polar coordinate system is obtained.\citep{Rudenberg1951}
Thus, by convention, the value of the practical infinity $r_{\infty}$
is typically chosen in fully numerical diatomic calculations in the
same way as in atomic calculations, that is, by specifying the radius
of a large sphere centered at the origin, which encloses the system.
The corresponding $\mu$ value can then be obtained as
\begin{align}
\mu_{\infty}= & \text{arcosh}\frac{r_{\infty}}{R_{h}}=\text{arcosh}\frac{2r_{\infty}}{R}.\label{eq:mumax}
\end{align}
Alternatively, since $\cosh\mu$ is large while $\mu$ is by definition
non-negative, one can furthermore approximate
\begin{align}
\cosh\mu= & \frac{1}{2}\left(e^{\mu}+e^{-\mu}\right)\approx\frac{1}{2}e^{\mu}\label{eq:coshmu-limit}
\end{align}
which yields a simpler approximate form
\begin{equation}
\mu_{\infty}\approx\log\frac{4r_{\infty}}{R}.\label{eq:mumax-limit}
\end{equation}

Calculations in the curvilinear coordinate system defined by \eqref{x-eq, y-eq, z-eq}
will require knowledge of the scale factors
\begin{align}
h_{i}(\xi,\eta,\phi)= & \sqrt{\left(\partial_{i}x\right)^{2}+\left(\partial_{i}y\right)^{2}+\left(\partial_{i}z\right)^{2}}.\label{eq:scalefac}
\end{align}
These are straightforwardly obtained as
\begin{align}
h_{\phi}= & R_{h}\sinh\mu\sin\nu,\label{eq:scale-phi}\\
h_{\nu}= & R_{h}\sqrt{\sinh^{2}\mu+\sin^{2}\nu},\label{eq:scale-nu}\\
h_{\mu}= & R_{h}\sqrt{\sinh^{2}\mu+\sin^{2}\nu}.\label{eq:scale-mu}
\end{align}
The volume element is then obtained as
\begin{align}
{\rm d}V= & h_{\phi}h_{\nu}h_{\mu}{\rm d}\phi{\rm d}\nu{\rm d}\mu\label{eq:dV-0}\\
= & R_{h}^{3}\sinh\mu\sin\nu\left(\sinh^{2}\mu+\sin^{2}\nu\right){\rm d}\phi{\rm d}\nu{\rm d}\mu\label{eq:dV-1}\\
= & R_{h}^{3}\sinh\mu\sin\nu\left(\cosh^{2}\mu-\cos^{2}\nu\right){\rm d}\phi{\rm d}\nu{\rm d}\mu.\label{eq:dV}
\end{align}
Identifying the angular element of the spherical polar coordinate
system

\begin{equation}
{\rm d}\Omega=\sin\nu{\rm d}\nu{\rm d}\phi,\label{eq:angfac}
\end{equation}
over which the spherical harmonics are orthonormal, the volume element
is obtained in the final form
\begin{align}
{\rm d}V= & R_{h}^{3}\sinh\mu\left(\cosh^{2}\mu-\cos^{2}\nu\right){\rm d}\mu{\rm d}\Omega.\label{eq:volel}
\end{align}
Comparing \eqref{r1, r2, volel}, it is seen that the volume element
contains the factor $r_{A}r_{B}$, which is the reason for the good
performance of the prolate spheroidal coordinate system. Nuclear attraction
integrals are smooth and have no singularities, as the $r_{A}^{-1}$
and $r_{B}^{-1}$ terms arising from the attraction of the two nuclei
are canceled by a factor in the volume element. In addition, as discussed
in \citeref{Lehtola2019c}, the presently used coordinate system is
especially nice for fully numerical approaches, as exponential functions
$\exp(-\zeta r_{A})$ or $\exp(-\zeta r_{B})$ turn to Gaussians in
the $(\mu,\nu,\phi)$ coordinates near the nuclei, thereby lacking
cups that would be difficult to represent numerically.

\subsection{Basis set\label{subsec:Basis-set}}

As in atoms, in diatomic molecules orbitals block by the $m$ quantum
number:\citep{McCullough1974}
\begin{equation}
\psi_{nm}(\boldsymbol{r})=\chi_{nm}(\mu,\nu)e^{im\phi}.\label{eq:diatom}
\end{equation}
Thus, a basis set is adopted in the form
\begin{align}
\chi_{nlm}(\mu,\nu,\phi)= & B_{n}(\mu)Y_{l}^{m}(\nu,\phi)\label{eq:basis}
\end{align}
where $Y_{l}^{m}$ are complex spherical harmonics, and $B_{n}(\mu)$
are one-dimensional finite element basis functions, which are commonly
known also as shape functions. The functions $B_{n}(\mu)$ are piece-wise
polynomials that are non-zero only within a single \emph{element}
$\mu\in[\mu_{\text{start}},\mu_{\text{end}}]$; see \citeref{Lehtola2019a}
for more information on the approach. In analogy to the atomic case
discussed in part I,\cite{Lehtola2019a} despite complex basis functions,
all the matrices in HF and DFT calculations end up being real --
even in the presence of a parallel magnetic field\citep{Lehtola2019d}
-- unless special approaches are used.\cite{Ostlund1972, Edwards1988, Lehtola2014a, Small2015a, Lehtola2016, Lehtola2016a}

The angular parts of matrix elements in the basis defined by \eqref{basis}
can be evaluated in closed form,\citep{McCullough1975,McCullough1986}
and most matrix elements will vanish by symmetry, as will be seen
later on in the manuscript. As in the atomic case, the same radial
grid is used for all angular momentum channels, as it greatly simplifies
the implementation; the total number of basis functions again being
given by the number of radial functions times that of angular functions.

\subsection{One-electron integrals\label{subsec:One-electron-integrals}}

As the volume element includes a $\cos^{2}\nu$ factor, the angular
basis set will not be orthonormal in contrast to the atomic case:
in addition to the diagonal coupling from $(l,m)$ to $(l,m)$, the
overlap matrix also includes couplings to $(l-2,m)$ and to $(l+2,m)$.
As angular integrals over cosines will appear here and there, we define
a cosine coupling coefficient as
\begin{equation}
\delta_{l_{1}l_{2}}^{(n)}=\int\left(Y_{l_{2}}^{0}\right)^{*}(\nu,\phi)Y_{l_{1}}^{0}(\nu,\phi)\cos^{n}\nu{\rm d}\Omega,\label{eq:coscpl}
\end{equation}
where the case $n=0$ yields the Kronecker delta symbol $\delta_{l_{1}l_{2}}$.
The cosine factors encountered in the present work can be expanded
in spherical harmonics as
\begin{align}
\cos\nu= & \frac{2}{3}\sqrt{3\pi}Y_{1}^{0}\label{eq:cos}\\
\cos^{2}\nu= & \frac{2}{3}\sqrt{\pi}Y_{0}^{0}+\frac{4}{15}\sqrt{5\pi}Y_{2}^{0}\label{eq:cos2}\\
\cos^{3}\nu= & \frac{2}{5}\sqrt{3\pi}Y_{1}^{0}+\frac{4}{35}\sqrt{7\pi}Y_{3}^{0}\label{eq:cos3}\\
\cos^{4}\nu= & \frac{2}{5}\sqrt{\pi}Y_{0}^{0}+\frac{8}{35}\sqrt{5\pi}Y_{2}^{0}+\frac{16}{105}\sqrt{\pi}Y_{4}^{0}\label{eq:cos4}\\
\cos^{5}\nu= & \frac{2}{7}\sqrt{3\pi}Y_{1}^{0}+\frac{8}{63}\sqrt{7\pi}Y_{3}^{0}+\frac{16}{693}\sqrt{11\pi}Y_{5}^{0}\label{eq:cos5}
\end{align}
and thus the values of $\delta_{l_{1}l_{2}}^{n}$ can be evaluated
easily from Gaunt coefficients
\begin{align}
Y_{l_{1}}^{m_{1}}(\Omega)Y_{l_{2}}^{m_{2}}(\Omega) & =\sum_{LM}G_{l_{1}l_{2},M}^{m_{1}m_{2},L}Y_{L}^{M}(\Omega)\label{eq:sph-closure}
\end{align}
as discussed in \citeref{Lehtola2011}. Note that we use an asymmetric
definition for the Gaunt coefficient in \eqref{sph-closure}, as discussed
in part I of the present series.\cite{Lehtola2019a}

All the necessary Gaunt coefficients are precomputed and stored in
memory at the start of the calculation. Note that unlike the atomic
case, in which the angular expansion is always limited, the angular
momentum $l$ may attain large values in the partial wave expansion
for diatomic molecules: for instance, the calculations on \ce{Cu2}
and \ce{CuLi} in the present work used expansions up to $l=46$.
Although elegant schemes for the sparse storage of Gaunt coefficient
tables have been discussed in the literature,\citep{Rasch2004,Pinchon2007}
in the present case only a small subset of $m$ values is needed --
from $m=0$ for $\sigma$ orbitals to $m=\pm3$ for $\varphi$ orbitals
-- and so a simple dense cubic array storage scheme $[(l_{1},m_{1}),(l_{2},m_{2});(L,M)]$
is sufficient for our work.

\subsubsection{Overlap\label{subsec:Overlap}}

Defining the radial integrals
\begin{align}
I_{ij}^{mn}= & \int B_{i}(\mu)B_{j}(\mu)\sinh^{m}\mu\cosh^{n}\mu{\rm d}\mu,\label{eq:radint}
\end{align}
the overlap integral can be written as
\begin{align}
S_{ij}= & \int B_{i}(\mu)Y_{l_{i}m_{i}}^{*}(\boldsymbol{\Omega})B_{j}(\mu)Y_{l_{j}m_{j}}(\boldsymbol{\Omega})R_{h}^{3}\sinh\mu\left(\cosh^{2}\mu-\cos^{2}\nu\right){\rm d}\mu{\rm d}\Omega\label{eq:overlap-0}\\
= & \left(R_{h}^{3}I_{ij}^{12}\delta_{l_{i},l_{j}}-R_{h}^{3}I_{ij}^{10}\delta_{l_{i},l_{j}}^{(2)}\right)\delta_{m_{i},m_{j}}.\label{eq:overlap}
\end{align}
The radial integrals are computed using Gauss--Chebyshev quadrature
as detailed in part I,\cite{Lehtola2019a} and integration over $\mu=[0,\infty)$
is again implied for brevity in the equations.

\subsubsection{Kinetic energy\label{subsec:Kinetic-energy}}

Also the kinetic energy is simple. The Laplacian is given by
\begin{align}
\nabla^{2}f= & \frac{1}{h_{\phi}h_{\nu}h_{\mu}}\left[\frac{\partial}{\partial\mu}\left(\frac{h_{\nu}h_{\phi}}{h_{\mu}}\frac{\partial f}{\partial\mu}\right)+\frac{\partial}{\partial\nu}\left(\frac{h_{\mu}h_{\phi}}{h_{\nu}}\frac{\partial f}{\partial\nu}\right)+\frac{\partial}{\partial\phi}\left(\frac{h_{\mu}h_{\nu}}{h_{\phi}}\frac{\partial f}{\partial\phi}\right)\right]\label{eq:lapl-def}\\
= & \frac{{\rm d}\phi{\rm d}\nu{\rm d}\mu}{{\rm d}V}R_{h}\Bigg[\sin\nu\frac{\partial}{\partial\mu}\left(\sinh\mu\frac{\partial f}{\partial\mu}\right)+\sinh\mu\frac{\partial}{\partial\nu}\left(\sin\nu\frac{\partial f}{\partial\nu}\right)\nonumber \\
+ & \frac{\sinh^{2}\mu+\sin^{2}\nu}{\sinh\mu\sin\nu}\frac{\partial}{\partial\phi}\left(\frac{\partial f}{\partial\phi}\right)\Bigg],\label{eq:lapl-start}\\
= & \frac{1}{R_{h}^{2}\left(\sinh^{2}\mu+\sin^{2}\nu\right)}\left[\frac{1}{\sinh\mu}\left(\frac{\partial}{\partial\mu}\left(\sinh\mu\frac{\partial f}{\partial\mu}\right)\right)+\frac{1}{\sin\nu}\frac{\partial}{\partial\nu}\left(\sin\nu\frac{\partial f}{\partial\nu}\right)\right]\nonumber \\
+ & \frac{1}{R_{h}^{2}\sinh^{2}\mu\sin^{2}\nu}\frac{\partial^{2}f}{\partial\phi^{2}}\label{eq:lapl-simpl}
\end{align}
in full agreement with Artemyev \emph{et al}.\citep{Artemyev2004}
Knowing that the spherical harmonics satisfy
\begin{align}
\frac{\partial^{2}}{\partial\phi^{2}}Y_{l}^{m}(\cos\nu,\phi)= & -m^{2}Y_{l}^{m},\label{eq:Ylm-m}\\
\left[\frac{1}{\sin\nu}\frac{\partial}{\partial\nu}\left(\sin\nu\frac{\partial}{\partial\nu}\right)-\frac{m^{2}}{\sin^{2}\nu}\right]Y_{l}^{m}\left(\cos\nu,\phi\right)= & -l(l+1)Y_{l}^{m}\left(\cos\nu,\phi\right),\label{eq:Ylm-l}
\end{align}
the Laplacian (\eqref{lapl-simpl}) of a basis function yields
\begin{align}
\nabla^{2}\chi_{j}= & \frac{1}{R_{h}^{2}\left(\sinh^{2}\mu+\sin^{2}\nu\right)}\Bigg[\frac{1}{\sinh\mu}\left(\frac{\partial}{\partial\mu}\left(\sinh\mu\frac{\partial B_{j}}{\partial\mu}\right)\right)Y_{l_{j}}^{m_{j}}\nonumber \\
- & B_{j}(\mu)\left(l_{j}(l_{j}+1)+\frac{m_{j}^{2}}{\sinh^{2}\nu}\right)Y_{l_{j}}^{m_{j}}\Bigg]\label{eq:laplchi}
\end{align}
in agreement with McCullough.\citep{McCullough1986} Thus, the kinetic
energy matrix element becomes 
\begin{align}
T_{ij}= & \int\chi_{i}^{*}(\boldsymbol{r})\left(-\frac{1}{2}\nabla^{2}\right)\chi_{j}(\boldsymbol{r}){\rm d}^{3}r\label{eq:T-0}\\
= & -\frac{1}{2}\int R_{h}B_{i}(\mu)\left(\frac{\partial}{\partial\mu}\left(\sinh\mu\frac{\partial B_{j}}{\partial\mu}\right)\right)\int\left(Y_{l_{i}}^{m_{i}}\right)^{*}Y_{l_{j}}^{m_{j}}{\rm d}\mu{\rm d}\Omega\nonumber \\
 & +\frac{1}{2}\int R_{h}B_{i}(\mu)\left[l_{j}(l_{j}+1)\sinh\mu+\frac{m_{j}^{2}}{\sinh\mu}\right]d\mu\int\left(Y_{l_{i}}^{m_{i}}\right)^{*}Y_{l_{j}}^{m_{j}}{\rm d}\Omega.\label{eq:T-1}
\end{align}
Finally, the first term can be symmetrized by invoking integration
by parts, as in the atomic case discussed in part I,\cite{Lehtola2019a}
yielding the kinetic energy matrix elements in the final form
\begin{equation}
T_{ij}=\frac{R_{h}}{2}\left[D_{i,j}+l_{j}(l_{j}+1)I_{1,2}^{10}+m_{j}^{2}I_{i,j}^{-1,0}\right]\delta_{l_{i},l_{j}}\delta_{m_{i},m_{j}},\label{eq:kinetic}
\end{equation}
where we have defined the radial integral 
\begin{align}
D_{1,2}= & \int\sinh\mu\frac{\partial B_{1}}{\partial\mu}\frac{\partial B_{2}}{\partial\mu}{\rm d}\mu.\label{eq:kin-D}
\end{align}

The examination of \eqref{T-1} shows that the kinetic energy density
diverges for $\mu\to0$ for $m\neq0$. This means that non-$\sigma$
states must vanish at $\mu=0$ 
\begin{equation}
\psi_{m}(\mu=0,\nu)=0,m\neq0.\label{eq:boundary}
\end{equation}
Unlike the atomic case discussed in part I,\cite{Lehtola2019a} the
used radial basis set must then depend on the value $m$. However,
\eqref{boundary} can be satisfied in the finite element implementation
as described in \citeref{Lehtola2019a} by removing the first shape
function of the first radial element for basis functions with $m\neq0$,
which is easily done in the C++ program.

\subsubsection{Nuclear attraction\label{subsec:Nuclear-attraction}}

As was stated above in \subsecref{Coordinate-system}, the nuclear
attraction integrals become easy for quadrature in the prolate spheroidal
coordinate system, as the singularities at the nuclei are canceled
out by factors in the volume element. The nuclear attraction integral
is
\begin{align}
V_{ij}= & \int\chi_{i}^{*}(\boldsymbol{r})\left(-\frac{Z_{A}}{r_{A}}-\frac{Z_{B}}{r_{B}}\right)\chi_{2}(\boldsymbol{r}){\rm d}^{3}r\label{eq:nuclear-0}\\
= & -R_{h}^{2}\int\chi_{i}^{*}(\boldsymbol{r})\left[\left(Z_{A}+Z_{B}\right)\cosh\mu+\left(Z_{B}-Z_{A}\right)\cos\nu\right]\chi_{j}(\boldsymbol{r})\sinh\mu{\rm d}\mu{\rm d}\Omega\label{eq:nuclear-1}
\end{align}
from which the integral is obtained in final form as
\begin{align}
V_{ij}= & -R_{h}^{2}\left(Z_{A}+Z_{B}\right)I_{i,j}^{11}\delta_{l_{i},l_{j}}\delta_{m_{i},m_{j}}-R_{h}^{2}\left(Z_{B}-Z_{A}\right)I_{i,j}^{10}\delta_{l_{i},l_{j}}^{(1)}\delta_{m_{i},m_{j}}.\label{eq:nuclear}
\end{align}

\subsubsection{Radial moments\label{subsec:Radial-moments}}

Radial moments of the density about the nuclei can be calculated using
\begin{align}
r_{A/B;ij}^{-1} & =R_{h}^{2}\left(I_{i,j}^{11}\delta_{l_{i},l_{j}}\mp I_{i,j}^{10}\delta_{l_{i},l_{j}}^{(1)}\right)\delta_{m_{i},m_{j}}\label{eq:rm1AB}
\end{align}
and
\begin{align}
r_{A/B} & =R_{h}(\cosh\mu\pm\cos\nu)\label{eq:r1AB0}\\
r_{A/B}^{2} & =R_{h}(\cosh^{2}\mu\pm2\cosh\mu\cos\nu+\cos^{2}\nu)\label{eq:r2AB0}\\
r_{A/B}^{3} & =R_{h}(\cosh^{3}\mu\pm3\cosh^{2}\mu\cos\nu+3\cosh\mu\cos^{2}\nu\pm\cos^{3}\nu)\label{eq:r3AB0}
\end{align}
from which
\begin{align}
\braket{r_{A/B}}_{ij}= & R_{h}^{4}\left[I_{i,j}^{13}\delta_{l_{i},l_{j}}\pm I_{i,j}^{12}\delta_{l_{i},l_{j}}^{(1)}-I_{i,j}^{11}\delta_{l_{i},l_{j}}^{(2)}\mp I_{i,j}^{10}\delta_{l_{i},l_{j}}^{(3)}\right]\delta_{m_{i},m_{j}}\label{eq:r1AB}\\
\braket{r_{A/B}}_{ij}^{2}= & R_{h}^{5}\left[I_{i,j}^{14}\delta_{l_{i},l_{j}}\pm2I_{i,j}^{13}\delta_{l_{i},l_{j}}^{(1)}\mp2I_{i,j}^{11}\delta_{l_{i},l_{j}}^{(3)}-I_{i,j}^{10}\delta_{l_{i},l_{j}}^{(4)}\right]\delta_{m_{i},m_{j}}\label{eq:r2AB}\\
\braket{r_{A/B}}_{ij}^{3}= & R_{h}^{6}\left[I_{i,j}^{15}\delta_{l_{i},l_{j}}\pm3I_{i,j}^{14}\delta_{l_{i},l_{j}}^{(1)}+2I_{i,j}^{13}\delta_{l_{i},l_{j}}^{(2)}\mp2I_{i,j}^{12}\delta_{l_{i},l_{j}}^{(3)}-3I_{i,j}^{11}\delta_{l_{i},l_{j}}^{(4)}\mp I_{i,j}^{10}\delta_{l_{i},l_{j}}^{(5)}\right]\delta_{m_{i},m_{j}}\label{eq:r3AB}
\end{align}
In \eqrangeref{rm1AB}{r3AB}, the upper sign corresponds to placing
the origin at the left-hand atom A at $z=-R_{h}$, while the lower
sign corresponds to placing the origin at the right-hand atom B at
$z=R_{h}$. The radial expectation value with respect to the origin
at the geometrical center of the molecule is
\begin{align}
\braket{r^{2}}_{ij}= & R_{h}^{5}\left[\left(I_{i,j}^{14}-I_{i,j}^{12}\right)\delta_{l_{i},l_{j}}+I_{i,j}^{10}\left(\delta_{l_{i},l_{j}}^{(2)}-\delta_{l_{i},l_{j}}^{(4)}\right)\right]\delta_{m_{i},m_{j}}.\label{eq:radexp}
\end{align}

\subsubsection{Electric field\label{subsec:Electric-field}}

The orbitals block by $m$ even in the presence of an electric field
parallel to the molecular bond, \emph{i.e. }in the $z$ direction;
the analogous case for magnetic fields is discussed in \citeref{Lehtola2019d}.
The $z$ component of the dipole operator is given by
\begin{align}
\mu_{z;ij}= & \int\chi_{i}^{*}(\boldsymbol{r})z\chi_{j}(\boldsymbol{r}){\rm d}V\label{eq:dip-0}\\
= & \int B_{i}(\mu)B_{j}(\mu)R_{h}\cosh\mu\cos\nu\cdot R_{h}^{3}\sinh\mu\left(\cosh^{2}\mu-\cos^{2}\nu\right){\rm d}\mu{\rm d}\Omega\left(Y_{l_{i}}^{m_{i}}\right)^{*}Y_{l_{j}}^{m_{j}}\label{eq:dip-1}\\
= & R_{h}^{4}\left[I_{i,j}^{13}\delta_{l_{i},l_{j}}^{(1)}-I_{i,j}^{11}\delta_{l_{i},l_{j}}^{(3)}\right]\delta_{m_{i},m_{j}}.\label{eq:dipmom}
\end{align}
The $zz$ component of the quadrupole operator is 
\begin{align}
\Theta_{zz}= & \frac{1}{2}\left(3z^{2}-r^{2}\right)=\frac{R_{h}^{2}}{2}\left[3\cosh^{2}\mu\cos^{2}\nu-\cosh^{2}\mu-\cos^{2}\nu+1\right]\label{eq:quadop}
\end{align}
which has the matrix element
\begin{align}
\Theta_{zz;ij}= & \int\chi_{i}^{*}(\boldsymbol{r})\Theta_{zz}\chi_{j}(\boldsymbol{r}){\rm d}^{3}r\label{eq:quadmom-def}\\
= & \frac{R_{h}^{5}}{2}\left[\left(I_{ij}^{12}-I_{ij}^{14}\right)\delta_{l_{i},l_{j}}+\left(3I_{ij}^{14}-I_{ij}^{10}\right)\delta_{l_{i},l_{j}}^{(2)}+\left(I_{ij}^{10}-3I_{ij}^{12}\right)\delta_{l_{i},l_{j}}^{(4)}\right]\delta_{m_{i},m_{j}}.\label{eq:quadmom}
\end{align}
The nuclear contributions to the electric dipole and quadrupole moments
are
\begin{align}
\mu_{z}^{\text{nuc}}= & R_{h}(Z_{2}-Z_{1}),\label{eq:dipnuc}\\
\Theta_{zz}^{\text{nuc}}= & R_{h}^{2}\left(Z_{1}+Z_{2}\right).\label{eq:quadnuc}
\end{align}
\Eqref{dip-0,quadnuc} are with respect to the origin; moments with
respect to other origins such as the center of mass or center of charge
are deferred to future work.

\subsection{Two-electron integrals\label{subsec:Two-electron-integrals}}

The two-electron integrals
\begin{align}
(ij|kl)= & \int\frac{\chi_{i}(\boldsymbol{r})\chi_{j}^{*}(\boldsymbol{r})\chi_{k}(\boldsymbol{r}')\chi_{l}^{*}(\boldsymbol{r}')}{\left|\boldsymbol{r}-\boldsymbol{r}'\right|}{\rm d}^{3}r{\rm d}^{3}r'\label{eq:tei-def}
\end{align}
can be readily evaluated with the help of the Neumann expansion, as
was originally pointed out by McCullough;\citep{McCullough1975,McCullough1986}
the same approach has also been used in \citerefs{Artemyev2004} and
\citenum{Zhang2015}, for example. As has been discussed by Ruedenberg
(equation 4.13 in reference \citenum{Rudenberg1951}), the Neumann
expansion of $r_{12}^{-1}$ is given by
\begin{align}
\frac{1}{r_{12}}= & \frac{4\pi}{R_{h}}\sum_{L=0}^{\infty}\sum_{M=-L}^{L}(-1)^{M}\frac{\left(L-|M|\right)!}{\left(L+|M|\right)!}P_{L}^{|M|}(\cosh\mu_{<})Q_{L}^{|M|}(\cosh\mu_{>})Y_{L}^{M}(\Omega_{1})\left(Y_{L}^{M}(\Omega_{2})\right)^{*},\label{eq:neumann}
\end{align}
where $P_{L}^{M}$, $Q_{L}^{M}$ are associated Legendre functions
of the first and second kind, respectively. Note that \eqref{neumann}
contains Legendre functions in two places: first, explicitly shown
with the argument $\cosh\mu\geq1$, and second, inside the spherical
harmonics with the more familiar branch $\left|\cos\nu\right|\leq1$.
The evaluation of the functions for the former case is not as well
known, but can be readily accomplished with software libraries described
in the literature.\citep{Gil1998,Schneider2010,Schneider2018} In
the present work, the library by Schneider and coworkers is used for
the evaluation of the Legendre functions in $\cosh\mu$.\citep{Schneider2010,Schneider2018} 

The Neumann expansion is analogous to the Laplace expansion that was
used for atomic calculations in part I,\cite{Lehtola2019a} with $P_{L}^{|M|}(\mu)$
taking the place of $r^{-L-1}$ in the large-radius integral, and
$Q_{L}^{|M|}$ taking the place of $r^{L}$ in the small-radius integral.
Analogously to the atomic case, $P_{L}^{|M|}(\cosh\mu)$ are regular
at $\mu=0$ but diverge as $\mu\to\infty$, while $Q_{L}^{|M|}(\cosh\mu)$
diverge at $\mu=0$ but go to zero for $\mu\to\infty$. Note, however,
that in contrast to the atomic case where the integrand only depends
on $L$, the diatomic integrals also depend on $M$, indicating a
more costly approach: even though the usual spherical polar coordinate
system is obtained by letting $R\to0$ in a diatomic calculation,\citep{Rudenberg1951}
the diatomic two-electron interactions still require more work than
the atomic calculations presented in part I.\cite{Lehtola2019a}

Substituting the Neumann expansion (\eqref{neumann}) into \eqref{tei-def}
yields
\begin{align}
(ij|kl)= & 4\pi R_{h}^{5}\sum_{L=0}^{\infty}\sum_{M=-L}^{L}(-1)^{M}\frac{\left(L-\left|M\right|\right)!}{\left(L+\left|M\right|\right)!}\int{\rm d}\mu_{1}{\rm d}\mu_{2}{\rm d}\Omega_{1}{\rm d}\Omega_{2}P_{L}^{\left|M\right|}(\cosh\mu_{<})Q_{L}^{\left|M\right|}(\cosh\mu_{>})\nonumber \\
 & \int\left(\cosh^{2}\mu_{1}-\cos^{2}\nu_{1}\right)\sinh\mu_{1}B_{i}(\mu_{1})B_{j}(\mu_{1})Y_{l_{i}}^{m_{i}}\left(\Omega_{1}\right)\left(Y_{l_{j}}^{m_{j}}\left(\Omega_{1}\right)\right)^{*}Y_{L}^{M}\left(\Omega_{1}\right)\nonumber \\
 & \int\left(\cosh^{2}\mu_{2}-\cos^{2}\nu_{2}\right)\sinh\mu_{2}B_{k}(\mu_{2})B_{l}(\mu_{2})Y_{l_{k}}^{m_{k}}(\Omega_{2})\left(Y_{l_{l}}^{m_{l}}(\Omega_{2})\right)^{*}\left(Y_{L}^{M}(\Omega_{2})\right)^{*}.\label{eq:tei0}
\end{align}
From here, we see that we must have
\begin{equation}
m_{j}-m_{i}=M=m_{k}-m_{l}\label{eq:M-cond}
\end{equation}
in order for the integral to be non-zero; the very same condition
was obtained also in the atomic case in \citeref{Lehtola2019a}. Furthermore,
the angular momentum algebra places limits on $L$ as in the atomic
case as 
\begin{equation}
L_{\min}\leq L\leq L_{\max}.\label{eq:L-cond}
\end{equation}
However, the cosine factors in \eqref{tei0} extend the range of the
coupled angular momentum by two in each direction compared to the
atomic case, yielding
\begin{align}
L_{\min}= & \max\{\left|l_{i}-l_{j}\right|,\left|l_{k}-l_{l}\right|\}-2,\label{eq:Lmin}\\
L_{\max}= & \min\{\left|l_{i}+l_{j}\right|,\left|l_{k}+l_{l}\right|\}+2,\label{eq:Lmax}
\end{align}
again signifiying more work than in an atomic calculation. The final
condition for the integral to be nonzero is that $Y_{L}^{M}$ must
exist, which gives
\begin{equation}
L_{\min}\geq|M|.\label{eq:LM-cond}
\end{equation}
\Eqrangeref{M-cond}{LM-cond} truncate the infinite sum in \eqref{tei0}
to a finite number of terms:
\begin{align}
(ij|kl)=\sum_{L=L_{\min}}^{L_{\max}} & \Bigg[I_{ij,kl}^{22,L|M|}G_{Ll_{i},m_{j}}^{Mm_{i},l_{j}}G_{Ll_{l},m_{k}}^{Mm_{l},l_{k}}-I_{ij,kl}^{02,L|M|}\tilde{G}_{Ll_{i},m_{j}}^{Mm_{i},l_{j}}G_{Ll_{l},m_{k}}^{Mm_{l},l_{k}}\nonumber \\
 & -I_{ij,kl}^{20,L|M|}G_{Ll_{i},m_{j}}^{Mm_{i},l_{j}}\tilde{G}_{Ll_{l},m_{k}}^{Mm_{l},l_{k}}+I_{ij,kl}^{00,L|M|}\tilde{G}_{Ll_{i},m_{j}}^{Mm_{i},l_{j}}\tilde{G}_{Ll_{l},m_{k}}^{Mm_{l},l_{k}}\Bigg],\label{eq:tei-final}
\end{align}
where $M$, $L_{\min}$, $L_{\max}$ and are given by \eqref{M-cond, Lmin, Lmax},
respectively, $I_{ij,kl}^{\alpha\beta,L|M|}$ are primitive integrals,
and we have defined a modified Gaunt coefficient as

\begin{equation}
\tilde{G}_{Ll_{i},m_{j}}^{Mm_{i},l_{j}}=\int\cos^{2}\nu Y_{l_{i}}^{m_{i}}\left(\Omega\right)\left(Y_{l_{j}}^{m_{j}}\right)^{*}\left(\Omega\right)Y_{L}^{M}\left(\Omega\right){\rm d}\Omega\label{eq:modgaunt-def}
\end{equation}
to account for the $\cos^{2}\nu$ terms. By employing \eqref{cos2, sph-closure},
the modified Gaunt coefficient in \eqref{modgaunt-def} can be written
in terms of the usual Gaunt coefficients as
\begin{equation}
\tilde{G}_{Ll_{i},m_{j}}^{Mm_{i},l_{j}}=\frac{2\sqrt{\pi}}{3}G_{L0,M}^{M0,L}G_{Ll_{i},m_{j}}^{Mm_{i},l_{j}}+\frac{4}{15}\sqrt{5\pi}\sum_{L'=L-2}^{L+2}G_{L2,M}^{M0,L'}G_{L'l_{i},m_{j}}^{Mm_{i},l_{j}}.\label{eq:modgaunt}
\end{equation}

\subsubsection{Primitive integrals\label{subsec:Primitive-integrals}}

The primitive integrals used in \eqref{tei-final} are defined as
\begin{align}
I_{ij,kl}^{\alpha\beta,L|M|}= & 4\pi R_{h}^{5}(-1)^{|M|}\frac{\left(L-\left|M\right|\right)!}{\left(L+\left|M\right|\right)!}\int\cosh^{\alpha}\mu_{1}\sinh\mu_{1}\cosh^{\beta}\mu_{2}\sinh\mu_{2}\nonumber \\
\times & B_{i}(\mu_{1})B_{j}(\mu_{1})B_{k}(\mu_{2})B_{l}(\mu_{2})P_{L}^{|M|}(\cosh\mu_{<})Q_{L}^{|M|}(\cosh\mu_{>}){\rm d}\mu_{1}{\rm d}\mu_{2}\label{eq:primint}
\end{align}
which alike the atomic case can be specialized into two cases: one
where all four functions are within the same element, and another
where $i$ and $j$ are in one element and $k$ and $l$ are in another.
Note that the expression corresponding to \eqref{primint} of Artemyev
\emph{et al.} (equation 32 in reference \citenum{Artemyev2004}) is
missing the $\cosh^{\alpha}\mu_{1}\sinh\mu_{1}\cosh^{\beta}\mu_{2}\sinh\mu_{2}$
factors arising from the volume elements.

As can be seen from \eqref{tei-final}, four sets of primitive integrals
corresponding to $(\alpha\beta)=(00)$, $(02)$, $(20)$, $(22)$
are needed, again increasing the amount of work compared to an atomic
calculation, with the possible values for $L$ and $M$ ranging from
$L=0,\dots,2\left(l_{\max}+1\right)$ and $|M|=0,\dots,2m_{\max}$,
where $l_{\max}$ and $m_{\max}$ are the largest values of $l$ and
$m$ in the basis set.

Like in the atomic case, most of the two-electron integrals in big
calculations arise from interelement integrals, which are written
in the factorizable form
\begin{align}
I_{ij,kl}^{\alpha\beta,L|M|}= & 4\pi R_{h}^{5}(-1)^{|M|}\frac{\left(L-\left|M\right|\right)!}{\left(L+\left|M\right|\right)!}\left[\int_{\mu_{1}^{\min}}^{\mu_{1}^{\max}}\cosh^{\alpha}\mu_{1}\sinh\mu_{1}Q_{L}^{|M|}(\cosh\mu_{1})B_{i}(\mu_{1})B_{j}(\mu_{1}){\rm d}\mu_{1}\right]\nonumber \\
 & \times\left[\int_{\mu_{2}^{\min}}^{\mu_{2}^{\max}}\cosh^{\beta}\mu_{2}\sinh\mu_{2}B_{k}(\mu_{2})B_{l}(\mu_{2})P_{L}^{|M|}(\cosh\mu_{2}){\rm d}\mu_{2}\right],\label{eq:tei-inter}
\end{align}
where we have assumed that the element containing $ij$ is farther
from the origin than the one containing $kl$. As in the atomic case,
the factorization in \eqref{tei-inter} can be used in the Coulomb
and exchange matrix algorithms. Also alike the atomic case, the intraelement
integrals are evaluated in three steps:
\begin{align}
\phi_{kl}^{\beta,L|M|}(\mu)= & \int_{0}^{\mu}{\rm d}\mu'\cosh^{\beta}\mu'\sinh\mu'B_{k}(\mu')B_{l}(\mu')P_{L}^{|M|}(\cosh\mu'),\label{eq:tei-phi}\\
i_{ij,kl}^{\alpha\beta,L|M|}= & \int_{0}^{\infty}{\rm d}\mu\cosh^{\alpha}\mu\sinh\mu B_{i}(\mu)B_{j}(\mu)Q_{L}^{|M|}(\cosh\mu)\phi_{kl}^{\beta,L|M|}(\mu),\label{eq:tei-i}\\
I_{ij,kl}^{\alpha\beta,L|M|}= & i_{ij,kl}^{\alpha\beta,L|M|}+i_{kl,ij}^{\beta\alpha,L|M|},\label{eq:tei-prim}
\end{align}
where \eqref{tei-phi} is computed in slices in analogy to the atomic
treatment. Note that in the last step, \emph{i.e. }in \eqref{tei-prim},
both $ij\leftrightarrow kl$ and $\alpha\leftrightarrow\beta$ are
interchanged.

\subsubsection{Coulomb matrix\label{subsec:Coulomb-matrix}}

The evaluation of Coulomb and exchange matrices can be sped up significantly
by employing the same techniques as in the atomic case of part I.\cite{Lehtola2019a}
This has also been recognized early on by McCullough.\citep{McCullough1986}
The Coulomb matrix is given by
\begin{equation}
J_{ij}=\sum_{kl}(ij|kl)P_{kl}.\label{eq:J-def}
\end{equation}
Substituting the expression for the two-electron integrals (\eqref{tei-final})
into \eqref{J-def} one obtains
\begin{align}
J_{ij}= & \sum_{L_{\min}}^{L_{\max}}\Bigg[I_{ij,kl}^{22,L|M|}P_{kl}G_{Ll_{i},m_{j}}^{Mm_{i},l_{j}}G_{Ll_{l},m_{k}}^{Mm_{l},l_{k}}-I_{ij,kl}^{02,L|M|}P_{kl}\tilde{G}_{Ll_{i},m_{j}}^{Mm_{i},l_{j}}G_{Ll_{l},m_{k}}^{Mm_{l},l_{k}}\nonumber \\
 & -I_{ij,kl}^{20,L|M|}P_{kl}G_{Ll_{i},m_{j}}^{Mm_{i},l_{j}}\tilde{G}_{Ll_{l},m_{k}}^{Mm_{l},l_{k}}+I_{ij,kl}^{00,L|M|}P_{kl}\tilde{G}_{Ll_{i},m_{j}}^{Mm_{i},l_{j}}\tilde{G}_{Ll_{l},m_{k}}^{Mm_{l},l_{k}}\Bigg]\label{eq:J}
\end{align}
where $M$ is defined by the constraint in \eqref{M-cond}, and $L_{\min}$
and $L_{\max}$ are defined by \eqrangeref{Lmin}{LM-cond}. Because
the primitive integrals $I_{ij,kl}^{\alpha\beta,L|M|}$ only depend
on the radial part and the compound index $L|M|$, one can form the
Coulomb matrix efficiently in three steps, analogously to the atomic
calculations discussed in part I.\cite{Lehtola2019a} The key here
is to form radial helper matrices by summing over the angular contributions
as 
\begin{align}
P_{kl}^{L|M|} & =\sum_{kl}G_{Ll_{l},m_{k}}^{Mm_{l},l_{k}}P_{kl},\label{eq:JP-1}\\
\tilde{P}_{kl}^{L|M|} & =\sum_{kl}\tilde{G}_{Ll_{l},m_{k}}^{Mm_{l},l_{k}}P_{kl},\label{eq:JP-2}
\end{align}
contract them with the primitive integrals to yield radial-only Coulomb
matrices
\begin{align}
J_{ij}^{L|M|} & =\sum_{kl}I_{ij,kl}^{22,L|M|}P_{kl}^{L|M|}-\sum_{kl}I_{ij,kl}^{20,L|M|}\tilde{P}_{kl}^{L|M|}\label{eq:JJ-1}\\
\tilde{J}_{ij}^{L|M|} & =\sum_{kl}I_{ij,kl}^{00,L|M|}\tilde{P}_{kl}^{L|M|}-\sum_{kl}I_{ij,kl}^{02,L|M|}P_{kl}^{L|M|}\label{eq:JJ-2}
\end{align}
and last, unroll the radial-only Coulomb matrices into the full Coulomb
matrix as
\begin{equation}
J_{ij}=J_{ij}^{L|M|}G_{Ll_{i},m_{j}}^{Mm_{i},l_{j}}+\tilde{J}_{ij}^{L|M|}\tilde{G}_{Ll_{i},m_{j}}^{Mm_{i},l_{j}}.\label{eq:J-unroll}
\end{equation}
The factorization of the interelement integrals can be used in \eqref{JJ-1,JJ-2}
to yield further speed improvements, whereas the contraction of intraelement
integrals can be done by matrix-vector multiplication.

\subsubsection{Exchange matrix\label{subsec:Exchange-matrix}}

For the exchange we have
\begin{equation}
K_{jk}^{\sigma}=\sum_{il}(ij|kl)P_{il}^{\sigma}\label{eq:K-def}
\end{equation}
which can also be made more efficient by summing over the angular
parts of $il$. However, while a single expansion over $L$ and $|M|$
sufficed for the Coulomb matrix, in the case of the exchange matrix,
the expansion has to be performed for all $jk$, making the calculation
significantly more expensive. As in the atomic case, the factorization
of the interelement integrals can be used to make the algorithm scale
better, while the intraelement contractions can be made faster by
storing a permuted set of the integrals in memory, allowing the use
of matrix-vector products.

\subsection{DFT\label{subsec:DFT}}

The implementation of DFT is exactly the same as in the atomic case
discussed in part I;\cite{Lehtola2019a} only the scale factors given
in \eqrangeref{scale-phi}{scale-mu} are different. As in the atomic
case, Gauss--Chebyshev quadrature is used in the $\nu$ direction,
whereas an equidistant grid is used for $\phi$. We have chosen $n_{\nu}=4l_{\max}+12$
and $n_{\phi}=4m_{\max}+5$ as the default values for diatomic calculations,
where the two extra points in the $\nu$ quadrature compared to the
atomic calculations have been added due to the $\cos^{2}\nu$ factor
in the volume element.

As was discussed for the atomic case in part I,\cite{Lehtola2019a}
popular functionals such as CAM-B3LYP;\citep{Yanai2004b} the Minnesota
functionals M11,\citep{Peverati2011b} N12-SX,\citep{Peverati2012a}
and MN12-SX;\citep{Peverati2012a} and the Head-Gordon group's $\omega$B97,\citep{Chai2008}
$\omega$B97X,\citep{Chai2008}, $\omega$B97X-V,\citep{Mardirossian2014a}
and $\omega$B97M-V\citep{Mardirossian2016} functionals employ a
range-decomposed Coulomb interaction\citep{Leininger1997}
\begin{equation}
\frac{1}{r_{12}}=\frac{\phi_{\text{sr}}(r_{12};\omega)}{r_{12}}+\frac{1-\phi_{\text{sr}}(r_{12};\omega)}{r_{12}},\label{eq:rangesep}
\end{equation}
in the exchange contribution, where the weight function is chosen
as
\begin{equation}
\phi_{\text{sr}}(r;\omega)=\text{erfc}(r;\omega).\label{eq:range-erfc}
\end{equation}
Implementing the aforementioned functionals in the present approach
would require a Neumann expansion for \eqref{rangesep, range-erfc}
alike \eqref{neumann}. As was mentioned in part I,\cite{Lehtola2019a}
we are not aware of suitable expansions even for the simpler atomic
case, but such expansions could be pursued in future work.

\subsection{Integral couplings\label{subsec:Integral-couplings}}

Having formulated expressions for all the integrals, it is useful
to study the couplings between the different angular blocks that have
been summarized in \tabref{One-electron-integral-couplings.}. In
contrast to the atomic case, where all one-electron operators were
diagonal in $l$ in the absence of electric fields, now only the kinetic
energy is diagonal, while everything else contains couplings between
different $l$ blocks in the basis set.

The wave function for one-electron systems in the absence of electromagnetic
fields is determined by the overlap, kinetic and nuclear attraction
matrices. Although the overlap and nuclear attraction matrices contain
couplings between the various $l$ channels in the basis set, their
matrix elements are independent of $l$: the $(l=0,m;l=0,m)$ block
has the same elements as the $(l=2,m;l=2,m)$ block. Also the dipole
and quadrupole matrix elements are independent of $l$. 

In contrast, while the kinetic energy operator does not couple different
values of $l$, its matrix elements are $l$-dependent, higher $l$
blocks carrying higher kinetic energy, as seen from \eqref{kinetic}.
This can be understood by the form of the basis set: higher values
of $l$ correspond to variation at finer scales, which carry higher
kinetic energy. Although the two-electron integrals also carry dependence
on the angular momentum, this analysis shows that the convergence
of a calculation is mainly determined by the kinetic energy.

\begin{table}
\begin{centering}
\begin{tabular}{lll}
Matrix type & Value of $l'$ coupling to $l$ & Radial elements dependent on $l$\tabularnewline
\midrule
\midrule 
Kinetic & $l$ & yes\tabularnewline
Nuclear attraction & $l$, $l\pm1$ & no\tabularnewline
Overlap & $l$, $l\pm2$ & no\tabularnewline
Dipole & $l\pm1$, $l\pm3$ & no\tabularnewline
Quadrupole & $l$, $l\pm2$, $l\pm4$ & no\tabularnewline
\end{tabular}
\par\end{centering}
\caption{One-electron integral couplings.\label{tab:One-electron-integral-couplings.}}
\end{table}

\subsection{Choice of basis set\label{subsec:Choice-of-basis}}

The CBS limit can be achieved in principle by systematically expanding
the basis set towards larger and larger values of $l$ for all the
values of $m$ included in the basis set. In addition to the angular
basis set, one must also converge the radial basis set, which implies
another truncation parameter. Although it is very well possible to
converge a calculation to the basis set limit by running a large number
of SCF calculations at increasing numbers of partial waves and radial
elements, this approach quickly becomes laborious, not to mention
overtly costly for heavier systems. Furthermore, as the coordinate
system depends on the bond length, in principle the basis set should
be converged separately at each geometry for every system.

The geometry dependence of the basis set requirements is easy to understand
by the following argument. Take a diatomic molecule with nuclei $Z_{A}$
and $Z_{B}$ separated by $R$. As has been discussed above and in
\citeref{Rudenberg1951}, letting $R\to0$, the prolate spheroidal
coordinate system approaches the spherical polar coordinate system.
This means that the diatomic calculation will approach an atomic one
for the compound nucleus $Z=Z_{A}+Z_{B}$. The angular expansion in
atomic calculations is extremely compact, as discussed in the first
part of this series:\cite{Lehtola2019a} for instance, the exact HF
ground state of all atoms from hydrogen to calcium is achievable with
just four angular functions -- $Y_{0}^{0}$, $Y_{1}^{-1}$, $Y_{1}^{0}$,
and $Y_{1}^{+1}$ -- barring symmetry breaking effects. However,
while the angular expansion becomes more compact when $R\to0$, a
heavier atom is also obtained at this limit, meaning that a larger
radial grid must be employed. Thus, in order to reproduce potential
energy surfaces, the radial grid should be converged at the \emph{smallest}
internuclear distance, whereas the angular grid should be converged
at the \emph{largest} internuclear distance.

Still, perhaps the worst feature of the naïve approach of running
SCF calculations with larger and larger basis sets is the lack of
estimates for the accuracy of any single calculation. The utility
of the present approach would be greatly increased were there a way
to easily choose a basis set for a given nuclear geometry with some
degree of control over the resulting accuracy. The radial and the
angular grids should be chosen in as balanced a way as possible to
yield the best possible accuracy with the least number of basis functions,
while minimizing the number of costly SCF calculations.

It is easy to see that the partial wave expansion should \textbf{not}
be the same for all values of $m$: the deepest and most compact orbitals
are the $1s$ core orbitals, which yield $\sigma$ orbitals in the
diatomic case. Atomic $p$ orbitals yield two $\pi$ orbitals ($m=\pm1)$
and one $\sigma$ orbital; atomic $d$ orbitals yield two $\delta$
($m=\pm2$) and $\pi$ ($m=\pm1$) orbitals and one $\sigma$ orbital;
and atomic $f$ orbitals yield two $\varphi$ ($m=\pm3$), $\delta$
($m=\pm2$), and $\pi$ ($m=\pm1$) orbitals and one $\sigma$ orbital,
all of which are less compact the $1s$ $\sigma$ orbital. Because
higher values of $l$ correspond to finer spatial resolution, it is
obvious that the number of partial waves should be highest for the
$\sigma$ orbitals, and decrease in increasing $|m|$; this is also
evident from the divergent $m^{2}/\sinh\mu$ term in the kinetic energy,
see \eqref{T-1,kinetic}. Thus, significant savings in the necessary
number of basis functions can be expected by the use of a non-uniform
angular grid, at no cost to the accuracy of the calculation. Unfortunately,
decoupling the number of partial waves in every $|m|$ channel introduces
further parameters that need to be optimized, as instead of a global
cutoff value $l_{\max}$ one now has to optimize the partial wave
cutoffs $l_{|m|}$ for $|m|=\sigma,\pi,\delta,\varphi$ in unison.

It is imaginable that an adaptive approach could be formulated for
the choice of the basis set. The determination of the radial grid
would be analogous to the atomic case, for which for both \emph{h}-adaptive\citep{Romanowski2009}
and \emph{p}-adaptive\citep{Flores1989,Flores1989a} approaches have
been presented in the literature, while the sufficiency of the angular
grid could be determined by determining the orbital gradient for rotations
into angular functions not included in the current basis set. However,
unless the occupations in each $|m|$ channel are predetermined, the
use of the Aufbau principle in the SCF calculation may result in incorrect
occupations if the angular basis set $l_{|m|}$ is unbalanced. For
instance, the use of an insufficiently large $l_{\pi}$ value may
result in $\pi$ orbital energies that are much too high, leading
to $\sigma$ orbitals being occupied by the Aufbau principle, instead.
It is thus apparent that the adaptive basis set should be determined
for a preset number of orbitals in each $|m|$ channel, but this would
then require additional user interaction.

Instead of using an adaptive approach relying on SCF calculations
as described above, we have found a simple and elegant solution to
the problem of basis set selection via a single-particle proxy approach.
Because higher $l$ values express smaller and smaller details in
the wave function -- especially close to the nuclei -- it makes
sense to simply study the convergence of the wave function close to
the nucleus, which can be approximated by the one-electron part of
the Hamiltonian operator, \emph{i.e.} the core Hamiltonian. In analogy
to completeness-optimization,\citep{Manninen2006,Lehtola2015} both
the radial and the angular basis set can be determined for any system
at any geometry by studying the convergence of a proxy for the molecular
energy 
\begin{equation}
E^{\text{proxy}}=\sum_{i\text{ occ}}\epsilon_{i}^{\text{core}}\label{eq:Eproxy}
\end{equation}
upon the addition of more radial elements or partial waves, where
$\epsilon_{i}^{\text{core}}$ are the eigenvalues of the core Hamiltonian
$\boldsymbol{H}_{0}=\boldsymbol{T}+\boldsymbol{V}$. In order to maintain
a balanced description, the addition trials increase the number of
radial elements or partial waves by two, as in homonuclear systems
the orbitals block by gerade/ungerade parity, which correspond to
even/odd-numbered partial waves. As orbitals for $m=\pm|m|$ are fully
degenerate for the core Hamiltonian, it suffices to only consider
the states with $m\geq0$ in the optimization.

The proxy corresponds to the one-electron part of the HF or DFT energy,
and differs from the full energy by the interactions of the electrons.
These omitted interactions extend the orbitals near the core, implying
that the proxy may overestimate the necessary number of partial waves
for the description of the wave function at an estimated accuracy
$\Delta$. Note that although hydrogenic orbitals (eigenfunctions
of $\boldsymbol{H}_{0}$) are notoriously bad for chemistry, as they
are typically both too compact due to the neglect of electronic repulsion
effects, as well as quickly become too diffuse to yield needed flexibility
in the molecular core and valence regions,\citep{Helgaker2000} this
is not a problem in the present approach as a predetermined $r_{\infty}$
poses limits on the diffuseness of the orbitals, and as the dimension
of the basis is not affected by the diagonalization of the core Hamiltonian.
For more discussion on the core guess and for an alternative one-electron
guess that could also be used as a proxy for basis set completeness,
see \citeref{Lehtola2019}.

Importantly, unlike an SCF based procedure for the adaptive formation
of the fully numerical basis set, the optimization of \eqref{Eproxy}
requires no solution of the SCF equations, and is thereby fast to
calculate; moreover, the $m$ channels are fully decoupled so the
$l_{|m|}$ values can be optimized separately. Furthermore, the number
of occupied orbitals in each $m$ channel can be chosen for any system
simply by considering the blocks of the periodic table in which the
nuclei in the calculation reside. Omitting the spin factor, in analogy
to our earlier work with completeness-optimized basis sets\citep{Lehtola2012a,Lehtola2013}
we choose the number of occupied proxy orbitals to cover the whole
block in the periodic table. That is, the number of occupied orbitals
in each $m$ channel is determined by counting the number of occupied
shells in the individual atom, and adding $1\sigma$ for $s$ shells,
$1\sigma1\pi$ for $p$ shells, $1\sigma1\pi1\delta$ for $d$ shells
and $1\sigma1\pi1\delta1\varphi$ for $f$ shells. For clarity, the
whole completeness-optimization procedure is shown step by step in
Algorithm \ref{algx:algorithm}.

Despite the differences between the proxy and the true HF / DFT energy,
we will demonstrate later in the manuscript that the error in the
SCF energy is similar to the estimate given by the proxy energy. To
our knowledge, this is the first time a non-uniform, truncated angular
basis set has been used in the literature. Note especially that McCullough's
approach\citep{McCullough1986} used an $l_{\max}$ value \emph{increasing}
in $m$ as $l_{m}=l_{\max}+|m|$ to maintain the same number of partial
waves in every $m$ channel, whereas our results show that a rapidly
decreasing $l_{m}$ is sufficient to yield fast convergence to the
CBS limit, bestowing significant speedups for the algorithm.

\begin{algorithm}
\begin{enumerate}
\item [0.] Specify the molecule ($Z_{A}$, $Z_{B}$, $R$) as well as the
wanted accuracy $\epsilon$.
\item Calculate the number of orbitals in each $m$ channel, based on the
blocks of the periodic table the atoms $Z_{A}$ and $Z_{B}$ belong
to, as described in the main text. For instance, Sc has 4 $s$ shells,
2 $p$ shells, and 1 $d$ shell; this means adding $4+2+1=7$ $\sigma$
orbitals, $3$ $\pi_{+1}$ and $\pi_{-1}$ orbitals, and $1$ $\delta_{+2}$
and $\delta_{-2}$ orbitals for a single Sc atom.
\item Initialize the radial and angular basis sets: set the number of radial
elements to $N_{\text{elem}}=1$, and the partial wave cutoff in each
channel to $l_{m}=|m|$.
\item Converge radial and angular basis sets for each channel $m\geq0$,
by looping down from the largest $|m|$ to 0.
\begin{enumerate}
\item Calculate the current value $E_{\text{cur}}$ for the proxy energy,
\eqref{Eproxy}, by solving the hydrogenic orbitals and energies for
the molecule ($Z_{A}$, $Z_{B}$, $R$) in the current fully numerical
basis set.\label{enu:Calculate-the-current}
\item Add more elements to the radial basis, $N_{\text{elem}}\to N_{\text{elem}}+2$,
and use \eqref{Eproxy} to calculate the resulting radial trial proxy
energy $E_{\text{rad}}$.\label{enu:Add-more-elements}
\item Add more partial waves to the $m$ channel, $l_{m}\to l_{m}+2$, and
use \eqref{Eproxy} to calculate the resulting angular trial proxy
energy $E_{\text{ang}}$.\label{enu:Add-more-partial}
\item Calculate the angular and radial energy lowerings $\Delta E_{\text{ang}}=E_{\text{cur}}-E_{\text{ang}}\geq0$
and $\Delta E_{\text{rad}}=E_{\text{rad}}-E_{\text{ang}}\geq0$.\label{enu:Calculate-energy-lowerings}
\item If $\Delta E_{\text{rad}}\geq\Delta E_{\text{ang}}$ and $\Delta E_{\text{rad}}\geq\epsilon$,
set $N_{\text{elem}}\to N_{\text{elem}}+2$ and go back to step \ref{enu:Calculate-the-current}.
\item If $\Delta E_{\text{ang}}>\Delta E_{\text{rad}}$ and $\Delta E_{\text{ang}}\geq\epsilon$,
set $l_{m}\to l_{m}+2$ and go back to step \ref{enu:Calculate-the-current}.
\item Otherwise, the wanted accuracy $\epsilon$ has been reached for the
$m$ channel; continue with the next $m$ value.
\end{enumerate}
\item Set the angular basis for $m=-|m|$ to that for $m=+|m|$: $l_{-|m|}=l_{|m|}$
for $m>0$.
\end{enumerate}
\caption{Formation of the diatomic basis set by the use of completeness-optimization
for the proxy energy.\label{algx:algorithm}}

\end{algorithm}

\section{Computational details\label{sec:Computational-details}}

The equations presented of \secref{Theory} have been implemented
in the C++ language. The \textsc{Armadillo} library has been used
for all matrix algebra,\citep{Sanderson2016,Sanderson2018} using
efficient basic linear algebra subroutine (\textsc{BLAS}) libraries
for the matrix operations. The program is parallellized with \textsc{OpenMP}
pragmas.

The one-electron and primitive two-electron integrals are precomputed
and stored in memory at the beginning of the calculation. $5N_{p}$
points are used for radial integrals; this should suffice even for
the highly non-linear integrals in DFT. The storage requirements for
the integrals are small, because only the auxiliary integrals are
stored instead of as instead of the full two-electron integral tensor.
Also, only the intraelement auxiliary integrals are stored as full
rank-4 tensors, while the interelement integrals are used in factorized
form that allows for faster formation of the Coulomb and exchange
matrices.

Exchange-correlation functionals are evaluated with the \textsc{Libxc}
library.\citep{Lehtola2018} Unless specified otherwise, SCF calculations
are initialized with the core guess, \emph{i.e.} eigenvectors of $\boldsymbol{H}_{0}=\boldsymbol{T}+\boldsymbol{V}$,
and the Aufbau principle is employed. A combination of the DIIS and
ADIIS accelerators is used in the SCF procedure.\citep{Pulay1980,Pulay1982,Hu2010}
Calculations are converged to an orbital gradient \emph{i.e.} DIIS
error of $10^{-7}$, unless otherwise stated.

Calculations within the fully spin-restricted, fully spin-unrestricted,
as well as spin-restricted open-shell via the constrained unrestricted
HF\citep{Tsuchimochi2010a,Tsuchimochi2011} formalisms are supported.
The orbitals obtained by full diagonalization of the Fock matrix by
$m$ block. As described in \citeref{Lehtola2019a}, because the finite
element basis set is never ill-conditioned, symmetric orthonormalization
is used to construct the molecular orbital basis set, but the basis
functions are normalized before the orthonormalization procedure.

Although the present implementation supports the same four grid types
discussed in part I for the atomic calculations,\citep{Lehtola2019a}
a linear grid in $\mu\in[0,\mu_{\infty}]$ is used for all calculations
in the present work, as the $(\mu,\nu,\phi)$ coordinate system already
yields wave functions that are smooth enough for efficient numerical
representation as was discussed in the Introduction. Following the
atomic calculations of \citeref{Lehtola2019a}, the calculations in
the present work employ $15^{\text{th}}$ order Lobatto elements in
the radial expansion, which allow for extremely fast convergence to
the radial basis set limit.

\section{Results\label{sec:Results}}

\subsection{HF limit energies\label{subsec:HF-limit-energies}}

\begin{sidewaystable*}
\begin{centering}
\begin{tabular}{llr@{\extracolsep{0pt}.}llllr@{\extracolsep{0pt}.}llllr@{\extracolsep{0pt}.}l}
molecule & bond length & \multicolumn{2}{c}{energy} &  & molecule & bond length & \multicolumn{2}{c}{energy} &  & molecule & bond length & \multicolumn{2}{c}{energy}\tabularnewline
\hline 
\hline 
$^1$\ce{CH+} & $2.137 a_0$ & -37&9099112 &  & $^3$\ce{NF} & $2.49 a_0$ & -153&8424212 &  & $^3$\ce{NCl} & $3.14 a_0$ & -513&9070135\tabularnewline
$^3$\ce{CH-} & $2.20 a_0$ & -38&2933200 &  & $^1$\ce{OF-} & $2.82 a_0$ & -174&2363416 &  & $^1$\ce{SiO} & $2.853 a_0$ & -363&8553418\tabularnewline
$^3$\ce{NH} & $1.9614 a_0$ & -54&9784239 &  & $^1$\ce{F2} & $2.668 a_0$ & -198&7734448 &  & $^3$\ce{PO-} & $2.90 a_0$ & -415&6564658\tabularnewline
$^1$\ce{OH-} & $1.781 a_0$ & -75&4188031 &  & $^2$\ce{F2-} & $3.52 a_0$ & -198&8623615 &  & $^3$\ce{SO} & $2.87 a_0$ & -472&3991048\tabularnewline
$^1$\ce{FH} & $1.7328 a_0$ & -100&0708025 &  & $^3$\ce{SiH-} & $2.94 a_0$ & -289&4646301 &  & $^1$\ce{SF-} & $3.22 a_0$ & -497&0283470\tabularnewline
$^1$\ce{C2} & $2.358 a_0$ & -75&4065652 &  & $^1$\ce{SH-} & $2.551 a_0$ & -398&1497909 &  & $^3$\ce{PF} & $3.015 a_0$ & -440&2339252\tabularnewline
$^2$\ce{CN} & 1.1718 Å & -92&2251382 &  & $^1$\ce{HCl} & $2.44 a_0$ & -460&1124493 &  & $^1$\ce{ClF} & $3.14 a_0$ & -558&9176263\tabularnewline
$^1$\ce{CN-} & $2.214 a_0$ & -92&3489506 &  & $^2$\ce{CP} & $3.08 a_0$ & -378&4746084 &  & $^1$\ce{SiS} & 1.93 Å & -686&5162842\tabularnewline
$^1$\ce{N2} & $2.068 a_0$ & -108&9938256 &  & $^1$\ce{CP-} & $3.00 a_0$ & -378&5615887 &  & $^1$\ce{P2} & $3.578 a_0$ & -681&5002553\tabularnewline
$^1$\ce{NO+} & $2.007 a_0$ & -128&9780515 &  & $^1$\ce{CS} & $2.89964 a_0$ & -435&3624203 &  & $^3$\ce{PS-} & $3.80 a_0$ & -738&3397074\tabularnewline
$^3$\ce{NO-}$^{a}$ & $2.36 a_0$ & -129&2801745 &  & $^2$\ce{SiN} & 1.575 Å & -343&2970269 &  & $^3$\ce{S2} & $3.642 a_0$ & -795&0915590\tabularnewline
$^1$\ce{CO} & $2.132 a_0$ & -112&7909072 &  & $^1$\ce{SiN-} & $2.94 a_0$ & -343&3623656 &  & $^1$\ce{SCl-} & $4.06 a_0$ & -857&1044186\tabularnewline
$^3$\ce{O2} & $2.270 a_0$ & -149&6687572 &  & $^1$\ce{NP} & $2.8173 a_0$ & -395&1883954 &  & $^1$\ce{Cl2} & $3.86 a_0$ & -919&0089345\tabularnewline
$^1$\ce{CF+} & $2.322 a_0$ & -136&9001348 &  & $^3$\ce{SN-} & $3.12 a_0$ & -451&9876493 &  & $^2$\ce{Cl2-} & $5.00 a_0$ & -919&0795637\tabularnewline
$^3$\ce{CF-} & $2.78 a_0$ & -137&2244562 &  &  &  & \multicolumn{2}{c}{} &  &  &  & \multicolumn{2}{c}{}\tabularnewline
\end{tabular}
\par\end{centering}
\caption{List of studied systems and restricted HF limit energies from \citeref{Jensen2005}.
All wave functions have $\Sigma$ symmetry.\protect \\
$^{a}$Reference \citeref{Jensen2005} erroneously reports a singlet
state for \ce{NO-}.\label{tab:List-MG}}
\end{sidewaystable*}

\begin{sidewaystable*}
\begin{centering}
\begin{tabular}{llr@{\extracolsep{0pt}.}llllr@{\extracolsep{0pt}.}llllr@{\extracolsep{0pt}.}l}
molecule & bond length & \multicolumn{2}{c}{energy} &  & molecule & bond length & \multicolumn{2}{c}{energy} &  & molecule & bond length & \multicolumn{2}{c}{energy}\tabularnewline
\hline 
\hline 
$^1$\ce{ScCl} & 2.229 Å & -1219&335786 &  & $^2$\ce{TiN} & 1.5802 Å & -902&769282 &  & $^1$\ce{NiC} & 1.631 Å & -1544&389546\tabularnewline
$^1$\ce{ScF} & 1.787 Å & -859&301233 &  & $^1$\ce{TiN+} & 1.586 Å & -902&526865 &  & $^1$\ce{NiSi} & 2.075 Å & -1795&561185\tabularnewline
$^1$\ce{ScH} & 1.775 Å & -760&277980 &  & $^3$\ce{VO-} & 1.615 Å & -1017&767081 &  & $^1$\ce{CuH} & 1.463 Å & -1639&514112\tabularnewline
$^2$\ce{ScN+} & 1.738 Å & -813&905803 &  & $^3$\ce{CrC} & 1.63 Å & -1080&868066 &  & $^1$\ce{CuF} & 1.745 Å & -1738&465275\tabularnewline
$^1$\ce{ScN} & 1.687 Å & -814&088437 &  & $^1$\ce{CrMn+} & 2.41 Å & -2192&467304$^{a}$ &  & $^1$\ce{CuCl} & 2.051 Å & -2098&548160\tabularnewline
$^1$\ce{ScO+} & 1.651 Å & -834&441524 &  & $^3$\ce{MnC-} & 1.615 Å & -1187&323508$^{a}$ &  & $^1$\ce{Cu2} & 2.22 Å & -3277&941606\tabularnewline
$^2$\ce{ScO} & 1.6682 Å & -834&674512 &  & $^3$\ce{FeC} & 1.670 Å & -1299&926272 &  & $^1$\ce{CuLi} & 2.26 Å & -1646&409856\tabularnewline
$^1$\ce{ScS-} & 2.188 Å & -1157&375086 &  & $^1$\ce{CoC-} & 1.564 Å & -1418&878845 &  & $^2$\ce{ZnH} & 1.595 Å & -1778&377824\tabularnewline
$^2$\ce{ScS} & 2.135 Å & -1157&338534 &  & $^3$\ce{CoO-} & 1.616 Å & -1456&143757 &  & $^2$\ce{ZnF} & 1.768 Å & -1877&344833\tabularnewline
\end{tabular}
\par\end{centering}
\caption{List of studied systems and restricted HF limit energies from \citeref{Williams2008}.
All wave functions have $\Sigma$ symmetry. $^{a}$The reference values
for \ce{CrMn+} and \ce{MnC-} are incorrect, see main text.\label{tab:List-TM}}
\end{sidewaystable*}

43 first- and second-row molecules from \citeref{Jensen2005}, and
27 transition metal molecules from \citeref{Williams2008} are studied;
the molecules and their ROHF limit energies are listed in \tabref{List-MG,List-TM},
respectively. All molecules have a wave function with $\Sigma$ symmetry;
that is, the net value for $m$ for the occupied orbitals is zero
in each spin channel.

The initial \textsc{HelFEM} calculations on \ce{NH}, \ce{ScF}, \ce{ScCl},
\ce{ScS}, \ce{TiN}, \ce{CrC}, \ce{MnC-}, \ce{FeC}, \ce{CrMn+},
and \ce{VO-} were found to converge to a higher-lying solution. In
most cases, it was enough to rectify the occupations of the initial
guess, but for \ce{ScS} the occupations had to be frozen for an additional
three iterations for the correct occupations to become stable in the
Aufbau solution. After a manual correction to the initial guess symmetry,
we were still unable to reproduce the energies reported in \citeref{Williams2008}
for \ce{CrC}, \ce{MnC-}, \ce{FeC}, and \ce{CrMn+}, requiring an
in-depth study of these systems.

We were unable to identify the proper ground state symmetry of these
four molecules with commonly-used Gaussian-basis programs, which lack
the possibility to enforce the full linear symmetry of the occupied
orbitals. The difficult convergence, especially in the case of \ce{MnC-},
is likely caused by a large number of low-lying configurations. Although
the symmetries can readily be restricted in \textsc{HelFEM}, the reliable
determination of the energy ranking of the various configurations
was found to require the use of large numerical basis sets: if an
insufficiently large numerical basis set was used, configurations
that are well-separated in energy in a fully converged calculation
erroneously turned out degenerate.

In order to allow quick exploratory calculations, linear symmetry
restrictions were implemented in the Gaussian-basis \textsc{Erkale}
program.\citep{erkale,Lehtola2012} Then, by enumerating all the possible
orbital occupations yielding $\Sigma$ symmetry, while restricting
the number of $\alpha$ orbitals in every $m$ channel be at least
that of $\beta$ orbitals, a brute-force search for the true ground
state configuration of \ce{CrC}, \ce{MnC-}, \ce{FeC}, and \ce{CrMn+}
was conducted in \textsc{Erkale} with the fully uncontracted aug-pc-$n$
basis sets,\citep{Jensen2001,Jensen2002b,Jensen2004,Jensen2013} yielding
the results in \tabref{Configuration-energy-differences}. In contrast
to the fully numerical calculations, configuration energy orderings
are reproduced correctly by even small Gaussian basis sets, as shown
by the data in \tabref{Configuration-energy-differences}.

According to the data in \tabref{Configuration-energy-differences},
the reference energy given for \ce{CrMn+} in \citeref{Williams2008}
is incorrect, as the absolute energy deviates by 2.6 eV from the reference
value of \citeref{Williams2008} even in the fully decontracted quintuple-$\zeta$
calculation; a similar deviation was also reproduced with the quintuple-$\zeta$
aug-cc-pV5Z correlation consistent basis set.\citep{Dunning1989,Kendall1992,Balabanov2005}
For this reason, \ce{CrMn+} was excluded from the present study.
The Gaussian basis set data also show that the ROHF limit energy given
in \citeref{Williams2008} for \ce{MnC-} is not fully converged,
as the variational calculation in the Gaussian basis reproduced a
lower energy than that of the fully numerical \textsc{x2dhf} calculation
of \citeref{Williams2008}.

Next, fixing the orbital occupations of \ce{CrC}, \ce{MnC-}, and
\ce{FeC} to the ground state configuration found with \textsc{Erkale},
it was found that the \textsc{HelFEM} calculations for \ce{CrC} and
\ce{MnC-} failed to converge within 50 SCF iterations with the core
guess. Further calculations in \textsc{Erkale} showed that the problem
was caused by the bad guess, which is especially poor for heavy atoms:\citep{Lehtola2019}
the Gaussian basis calculations also failed to converge with the core
guess. However, we have recently proposed a simple solution to this
problem in \citeref{Lehtola2019}: a good starting guess is obtained
simply by using radially screened nuclear charges, which can be obtained
from fully numerical calculations on atoms similar to the ones discussed
in part I.\citep{Lehtola2019a} The superposition of atomic potentials
(SAP) guess described in \citeref{Lehtola2019} was implemented in
\textsc{HelFEM}, and it was used with an LDA exchange-only potential
to form initial guesses for \ce{CrC}, \ce{MnC-}, and \ce{FeC},
after which the fully numerical calculations on these molecules converged
without problems. For \ce{MnC-} and \ce{CrMn+} at the studied geometries,
we find the fully numerical energies -1187.3240938 and -2192.3703627,
respectively. These are in good agreement with the values computed
with \textsc{Erkale} in the fully decontracted aug-pc-4 basis set:
-1187.3239603 and -2192.3700339, respectively. Our fully numerical
reference value for \ce{MnC-} is 586 $\mu E_{h}$ lower than the
one given in \citeref{Williams2008}, whereas the 2.64 eV discrepancy
for \ce{CrMn+} suggests that the ROHF limit value of \citeref{Williams2008}
does not correspond to the reported charge, spin state, and/or geometry.

\begin{table}
\begin{centering}
\begin{tabular}{lr@{\extracolsep{0pt}.}lr@{\extracolsep{0pt}.}lr@{\extracolsep{0pt}.}lr@{\extracolsep{0pt}.}lr@{\extracolsep{0pt}.}lr@{\extracolsep{0pt}.}lr@{\extracolsep{0pt}.}lr@{\extracolsep{0pt}.}l}
molecule & \multicolumn{4}{c}{un-aug-pc-1} & \multicolumn{4}{l}{un-aug-pc-2} & \multicolumn{4}{c}{un-aug-pc-3} & \multicolumn{4}{l}{un-aug-pc-4}\tabularnewline
 & \multicolumn{2}{c}{$\Delta$} & \multicolumn{2}{c}{$E-E_{\text{ref}}$} & \multicolumn{2}{c}{$\Delta$} & \multicolumn{2}{c}{$E-E_{\text{ref}}$} & \multicolumn{2}{c}{$\Delta$} & \multicolumn{2}{c}{$E-E_{\text{ref}}$} & \multicolumn{2}{c}{$\Delta$} & \multicolumn{2}{c}{$E-E_{\text{ref}}$}\tabularnewline
\hline 
\hline 
\ce{CrC} & 4&446 & 2&480 & 4&402 & 0&159 & 4&393 & 0&016 & 4&389 & 0&003\tabularnewline
\ce{MnC-} & 3&179 & 2&666 & 3&118 & 0&156 & 3&109 & 0&001 & 3&103 & -0&012\tabularnewline
\ce{FeC} & 1&677 & 2&787 & 1&542 & 0&169 & 1&514 & 0&015 & 1&509 & 0&002\tabularnewline
\ce{CrMn+} & 0&843 & 6&923 & 0&891 & 2&931 & 0&897 & 2&681 & 0&899 & 2&647\tabularnewline
\end{tabular}
\par\end{centering}
\caption{Energy difference $\Delta$ between the lowest configuration and second
lowest configuration for \ce{CrC}, \ce{MnC-}, \ce{FeC}, and \ce{CrMn+}
found in the brute-force search with the fully uncontracted aug-pc-$n$
basis sets (un-aug-pc-$n$), as well as the difference of the ground-state
energy from the Gaussian calculation to the numerical reference value
from \citeref{Williams2008}, $E-E_{\text{ref}}$. All values are
in eV.\label{tab:Configuration-energy-differences}}
\end{table}

\begin{figure}
\subfloat[$r_{\infty}=20a_{0}$]{\includegraphics[width=0.33\textwidth]{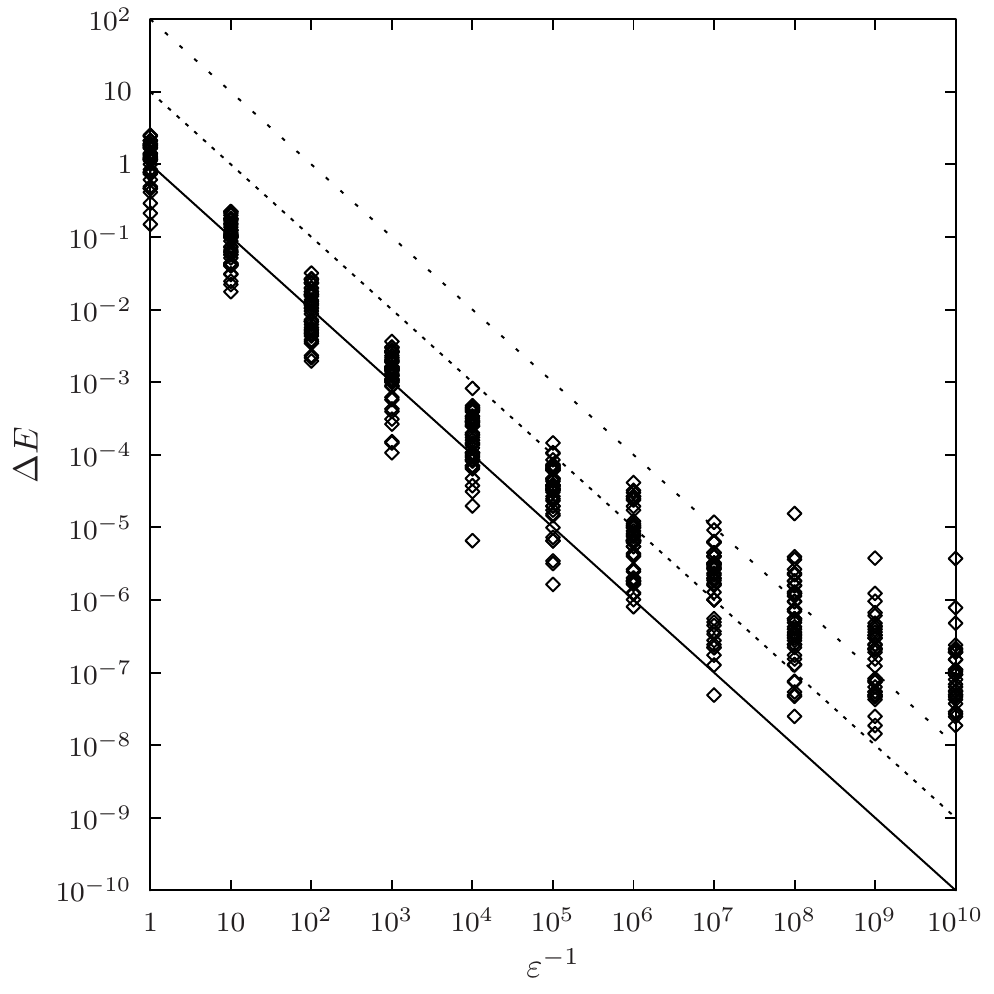}

}\subfloat[$r_{\infty}=40a_{0}$]{\includegraphics[width=0.33\textwidth]{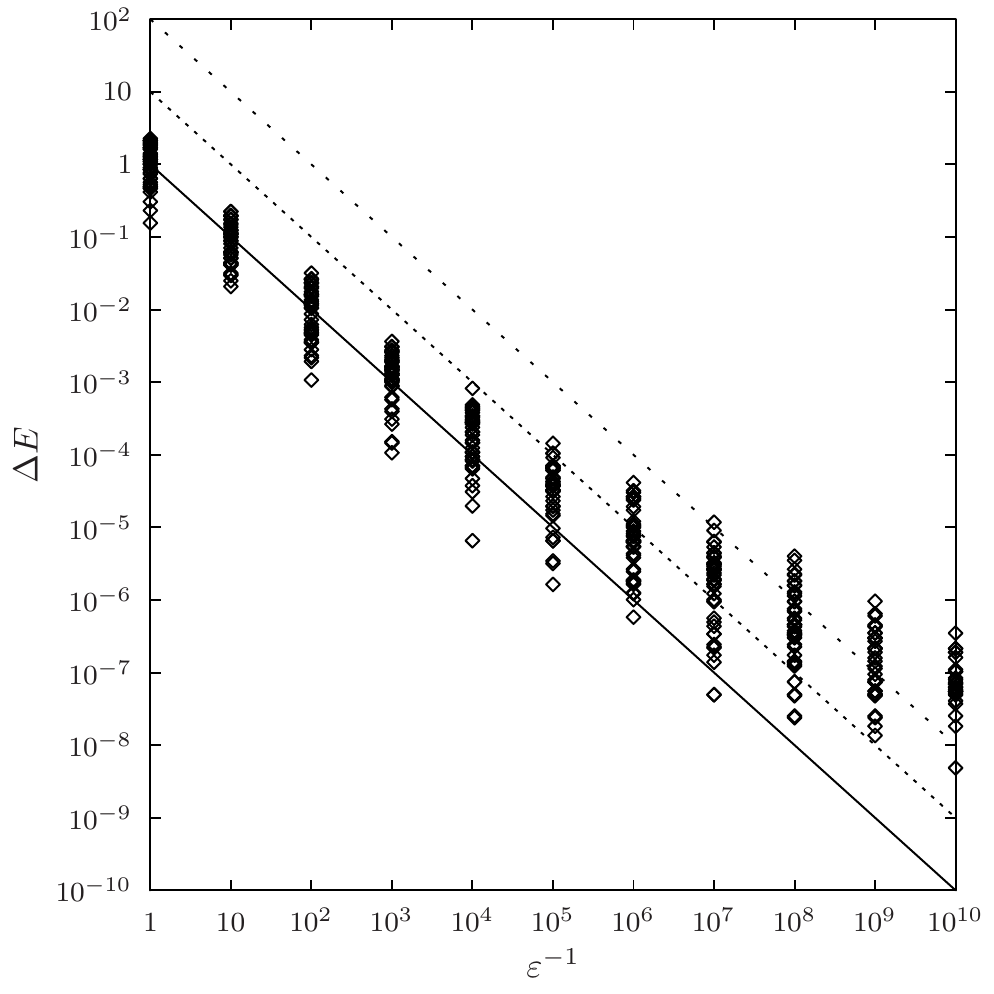}

}\subfloat[$r_{\infty}=60a_{0}$]{\includegraphics[width=0.33\textwidth]{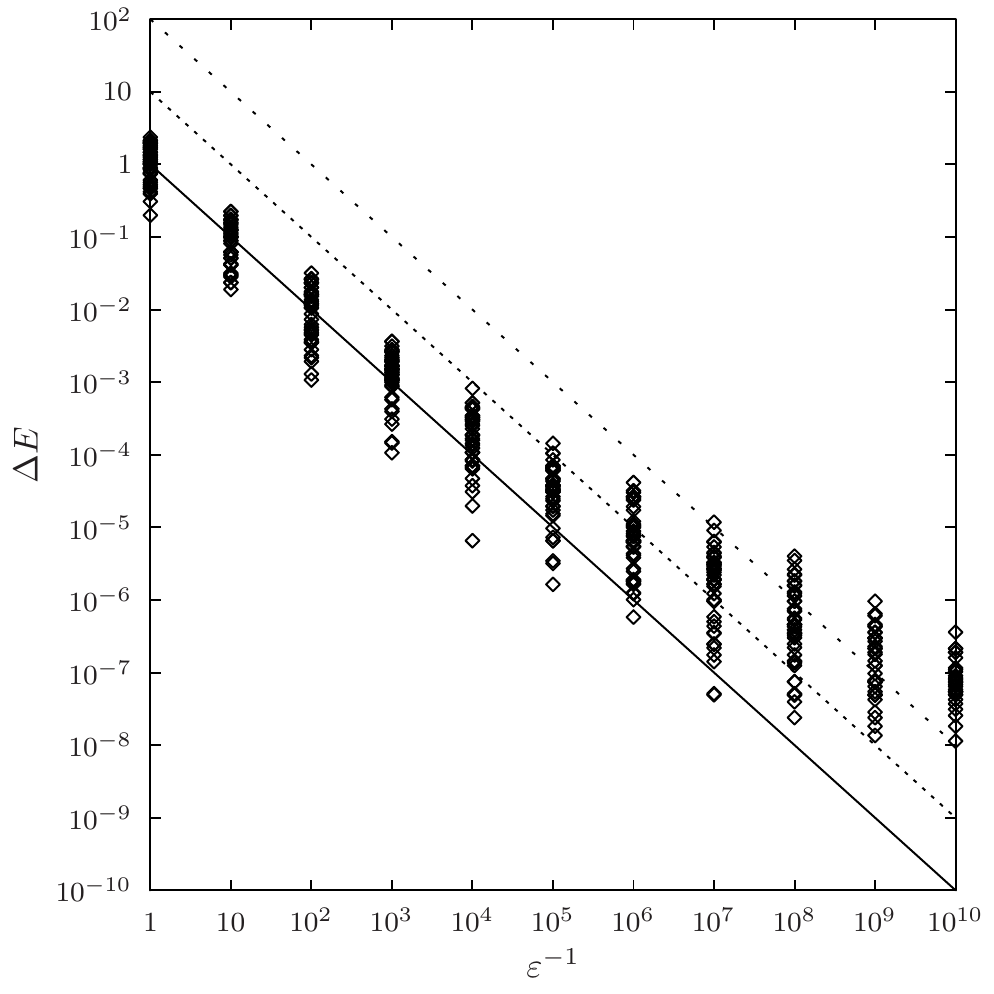}

}

\caption{Convergence of the restricted HF energy for 43 first- and second-row
molecules, compared to literature values from \citeref{Jensen2005}
given in \tabref{List-MG}. Note logarithmic scale. The ideal behavior
$\Delta E=\epsilon$ is represented by the solid line, with the dotted
and loosely dotted lines illustrating behavior corresponding to $\Delta E=10\epsilon$
and $\Delta E=100\epsilon$, correspondingly. \label{fig:conv-mg}}
\end{figure}

\begin{figure}
\subfloat[$r_{\infty}=20a_{0}$]{\includegraphics[width=0.33\textwidth]{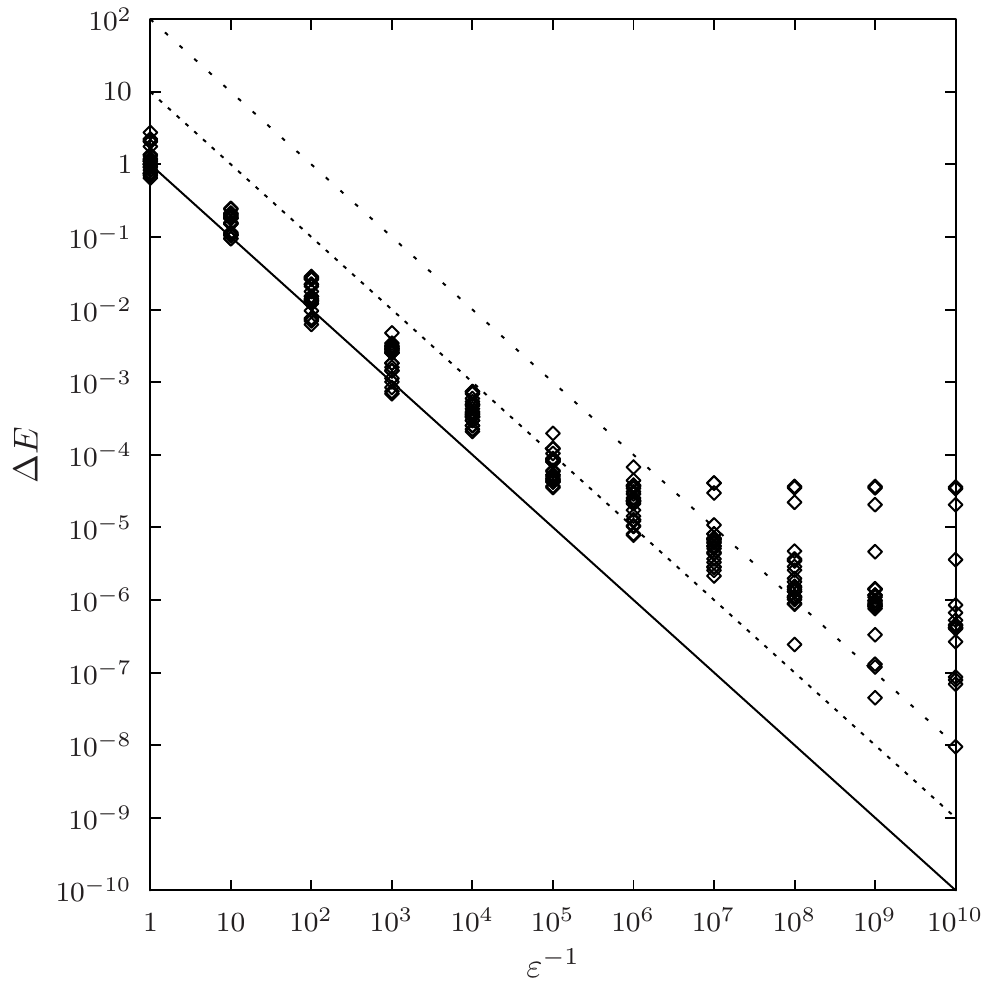}

}\subfloat[$r_{\infty}=40a_{0}$]{\includegraphics[width=0.33\textwidth]{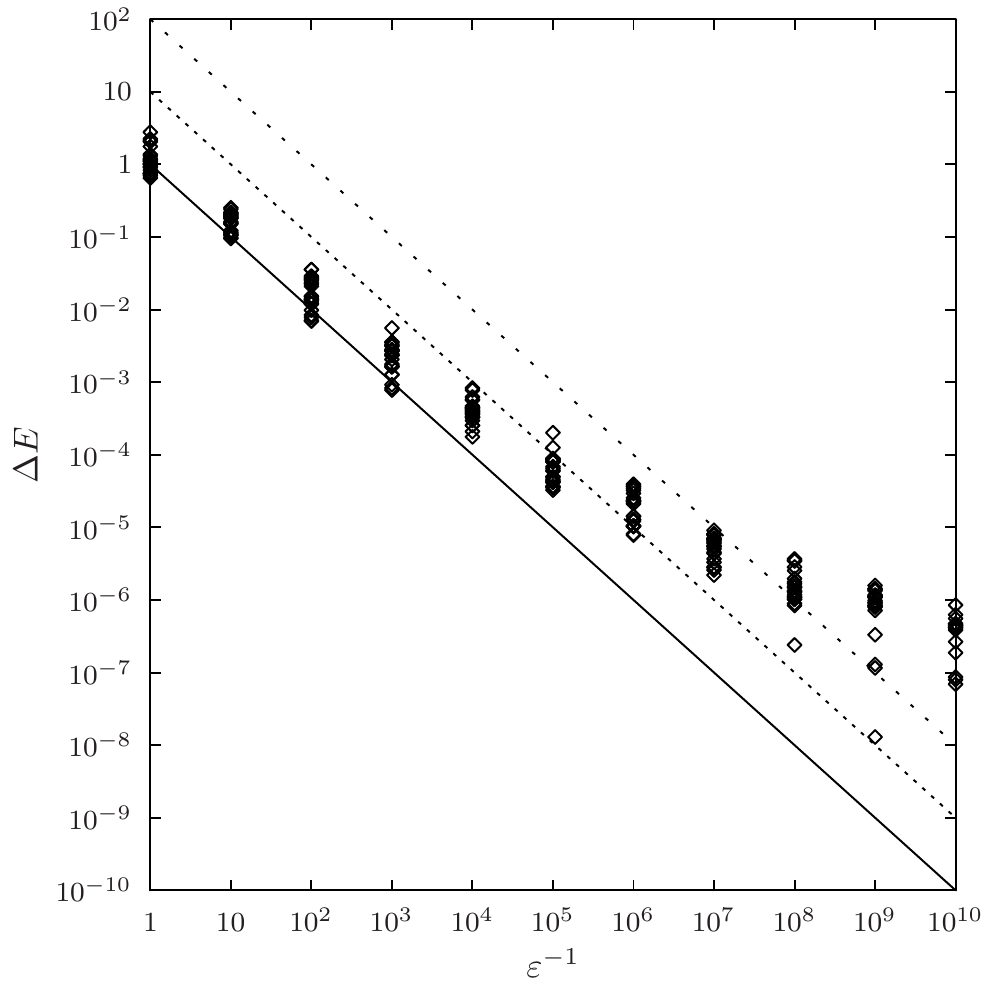}

}\subfloat[$r_{\infty}=60a_{0}$]{\includegraphics[width=0.33\textwidth]{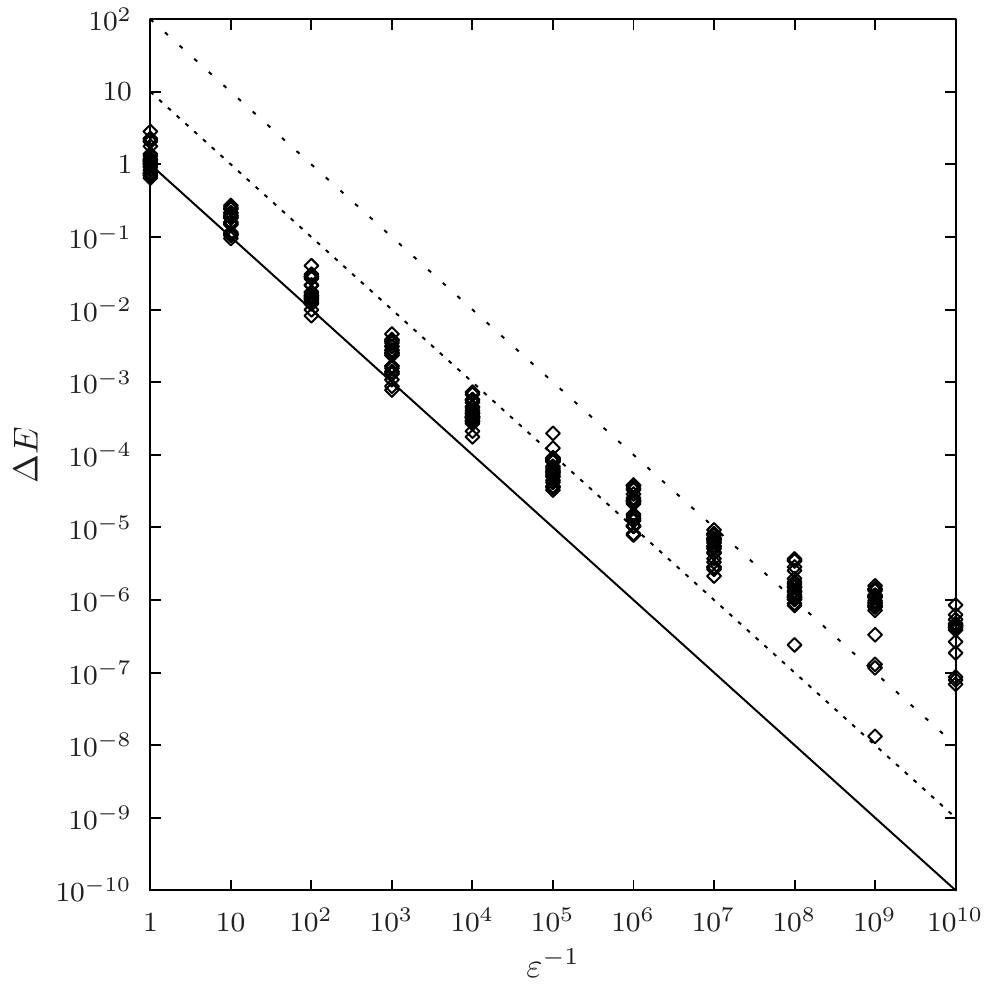}

}

\caption{Convergence of the restricted HF energy for 25 transition metal molecules,
compared to literature values from \citeref{Williams2008} given in
\tabref{List-TM}. \ce{MnC-} and \ce{CrMn+} have been excluded,
as the reference values are incorrect. Note logarithmic scale. The
notation is the same as in \figref{conv-mg}. \label{fig:conv-tm}}
\end{figure}

Excluding \ce{CrMn+} and \ce{MnC-} for which the literature values
are incorrect, we obtain the convergence behavior shown in \figref{conv-mg,conv-tm}
for the 43 main group and 25 transition metal molecules, respectively.
Both figures present results for $r_{\infty}=20a_{0}$, $r_{\infty}=40a_{0}$,
and $r_{\infty}=60a_{0}$. It is clear from these results that the
chosen proxy is remarkably successful in capturing the essential degrees
of freedom in the basis set, as the error in the self-consistent energy
is seen to follow that in the proxy within an order of magnitude until
$\epsilon=10^{-5}$, when the error starts to saturate. The error
levels off because of the finite accuracy of the reference data: the
HF limit energies for the main group and transition metal molecules
have been given with 7 and 6 decimals in \citerefs{Jensen2005} and
\citenum{Williams2008}, respectively, which were also repeated in
\tabref{List-MG,List-TM}.

As was discussed in the first part of this series dealing with atomic
calculations,\cite{Lehtola2019a} the value for the practical infinity
$r_{\infty}=20a_{0}$ is too small, yielding significant errors especially
in anionic systems, as can be seen from the outliers in \figref{conv-mg,conv-tm}
that do not exist in the $r_{\infty}=40a_{0}$ and $r_{\infty}=60a_{0}$
plots. Based on these results and those obtained for atoms in part
I,\cite{Lehtola2019a} we tentatively conclude that $r_{\infty}=40a_{0}$
should be sufficient for applications of the present method. However,
the convergence with respect to $r_{\infty}$ should be always checked,
as loosely bound anions be extremely diffuse -- especially in DFT
calculations -- as was discussed in the first part of the series.\cite{Lehtola2019a}

Despite the spread in the results and the finite accuracy of the reference
data, it is clear that choosing a basis with an estimated accuracy
of $\epsilon=10^{-10}$ should yield energies that are converged beyond
microhartree accuracy. Indeed, in some cases \textsc{HelFEM} reproduces
variational energies that are considerably \emph{lower} than the previous
reference values. For instance, the best energy reported for \ce{Cl2-}
in \citeref{Jensen2005} is $-919.0795637$, whereas the $\epsilon=10^{-10}$
basis set for $r_{\infty}=40a_{0}$ in \textsc{HelFEM} yields the
value $-919.0795645$ that is $0.8\mu E_{h}$ lower.

This is all the more striking, as the basis set used for the \textsc{HelFEM}
calculation is considerably smaller than the one that was used to
reproduce the value in \citeref{Jensen2005}. The more accurate \textsc{HelFEM}
value was obtained with $r_{\infty}=40a_{0}$, yielding 98 basis functions
in $\mu$. In contrast, the \textsc{x2dhf} calculation of \citeref{Jensen2005}
needed $r_{\infty}=400a_{0}$ and 535 points in the radial $\mu$
grid to converge the energy. As has been discussed above, \textsc{x2dhf}
uses an asymptotic expansion to calculate the Coulomb and exchange
potentials, requiring $r_{\infty}$ to be much larger than what it
would be based on the density alone, as in \textsc{HelFEM}. However,
since the size of the $\mu$ grid scales logarithmically in $r_{\infty}$
(\eqref{mumax-limit}), this is not a huge problem: $\mu_{\infty}\approx5.78$
in the \textsc{x2dhf} calculation, compared to $\mu_{\infty}\approx3.46$
in the \textsc{HelFEM} calculation, meaning that a 67\% larger $\mu$
grid would naïvely suffice in the \textsc{x2dhf} calculation.

Instead, the main reason \textsc{HelFEM} yields better accuracy despite
needing over five times fewer radial functions is that it uses high-order
elements that afford extremely fast convergence to the radial basis
set limit, as was found in \citeref{Lehtola2019a}. In addition to
a more compact radial expansion by over a factor of five, also the
angular expansion is much more compact in \textsc{HelFEM}: compared
to the 295 points in $\nu$ used for \ce{Cl2-} in \citeref{Jensen2005},
the more accurate \textsc{$r_{\infty}=40a_{0}$, $\epsilon=10^{-10}$
HelFEM} calculation in the present work employed only 39 $\sigma$
functions ($l_{\sigma}=38$) and 29 $\pi_{+1}$ and $\pi_{-1}$ functions
($l_{\pi}=29$), indicating a further reduction by an order of magnitude
in the size of the basis set.

\Tabref{Basis-MG, Basis-TM} show comparisons of the basis sets to
obtain results with a microhartree-level accuracy employed in \citerefs{Jensen2005}
and \citenum{Williams2008}, respectively, to the $r_{\infty}=40a_{0}$,
$\epsilon=10^{-10}$ basis sets used in the present work. These results
are fully in line with the discussion above for the case of \ce{Cl2-},
showing that the combination of high-order finite elements and the
partial wave expansion in \textsc{HelFEM} allow for basis sets that
are more compact by orders of magnitude compared to the approach used
in the \textsc{x2dhf} program.

\begin{sidewaystable*}
\begin{centering}
\begin{tabular}{l|lll|r@{\extracolsep{0pt}.}lr@{\extracolsep{0pt}.}lr@{\extracolsep{0pt}.}l||l|lll|r@{\extracolsep{0pt}.}lr@{\extracolsep{0pt}.}ll||l|lll|r@{\extracolsep{0pt}.}lr@{\extracolsep{0pt}.}lr@{\extracolsep{0pt}.}l}
 & \multicolumn{3}{c|}{\textsc{x2dhf}} & \multicolumn{6}{c||}{\textsc{HelFEM}} &  & \multicolumn{3}{c|}{\textsc{x2dhf}} & \multicolumn{5}{c||}{\textsc{HelFEM}} &  & \multicolumn{3}{c|}{\textsc{x2dhf}} & \multicolumn{6}{c}{\textsc{HelFEM}}\tabularnewline
molecule & $r_{\infty}$ & $n_{\mu}$ & $n_{\nu}$ & \multicolumn{2}{c}{$n_{\mu}$} & \multicolumn{2}{c}{$l_{\sigma}$} & \multicolumn{2}{c||}{$l_{\pi}$} & molecule & $r_{\infty}$ & $n_{\mu}$ & $n_{\nu}$ & \multicolumn{2}{c}{$n_{\mu}$} & \multicolumn{2}{c}{$l_{\sigma}$} & $l_{\pi}$ & molecule & $r_{\infty}$ & $n_{\mu}$ & $n_{\nu}$ & \multicolumn{2}{c}{$n_{\mu}$} & \multicolumn{2}{c}{$l_{\sigma}$} & \multicolumn{2}{c}{$l_{\pi}$}\tabularnewline
\hline 
\hline 
$^1$\ce{CH+} & 30 & 193 & 163 & \multicolumn{2}{c}{42} & \multicolumn{2}{c}{15} & \multicolumn{2}{c||}{11} & $^3$\ce{NF} & 200 & 319 & 175 & \multicolumn{2}{c}{70} & \multicolumn{2}{c}{19} & 15 & $^3$\ce{NCl} & 300 & 409 & 217 & \multicolumn{2}{c}{98} & \multicolumn{2}{c}{29} & \multicolumn{2}{c}{23}\tabularnewline
$^3$\ce{CH-} & 40 & 229 & 169 & \multicolumn{2}{c}{42} & \multicolumn{2}{c}{15} & \multicolumn{2}{c||}{11} & $^1$\ce{OF-} & 200 & 313 & 175 & \multicolumn{2}{c}{70} & \multicolumn{2}{c}{21} & 17 & $^1$\ce{SiO} & 200 & 313 & 175 & \multicolumn{2}{c}{70} & \multicolumn{2}{c}{25} & \multicolumn{2}{c}{19}\tabularnewline
$^3$\ce{NH} & 40 & 235 & 169 & \multicolumn{2}{c}{42} & \multicolumn{2}{c}{15} & \multicolumn{2}{c||}{13} & $^1$\ce{F2} & 300 & 337 & 175 & \multicolumn{2}{c}{70} & \multicolumn{2}{c}{21} & 15 & $^3$\ce{PO-} & 300 & 415 & 217 & \multicolumn{2}{c}{70} & \multicolumn{2}{c}{27} & \multicolumn{2}{c}{21}\tabularnewline
$^1$\ce{OH-} & 40 & 241 & 169 & \multicolumn{2}{c}{42} & \multicolumn{2}{c}{15} & \multicolumn{2}{c||}{13} & $^2$\ce{F2-} & 400 & 421 & 217 & \multicolumn{2}{c}{70} & \multicolumn{2}{c}{23} & 17 & $^3$\ce{SO} & 350 & 421 & 217 & \multicolumn{2}{c}{98} & \multicolumn{2}{c}{27} & \multicolumn{2}{c}{21}\tabularnewline
$^1$\ce{FH} & 30 & 193 & 169 & \multicolumn{2}{c}{70} & \multicolumn{2}{c}{17} & \multicolumn{2}{c||}{13} & $^3$\ce{SiH-} & 200 & 307 & 175 & \multicolumn{2}{c}{70} & \multicolumn{2}{c}{27} & 19 & $^1$\ce{SF-} & 300 & 403 & 217 & \multicolumn{2}{c}{98} & \multicolumn{2}{c}{29} & \multicolumn{2}{c}{23}\tabularnewline
$^1$\ce{C2} & 40 & 223 & 169 & \multicolumn{2}{c}{42} & \multicolumn{2}{c}{17} & \multicolumn{2}{c||}{11} & $^1$\ce{SH-} & 300 & 421 & 217 & \multicolumn{2}{c}{70} & \multicolumn{2}{c}{27} & 21 & $^3$\ce{PF} & 300 & 409 & 217 & \multicolumn{2}{c}{70} & \multicolumn{2}{c}{27} & \multicolumn{2}{c}{21}\tabularnewline
$^2$\ce{CN} & 40 & 229 & 169 & \multicolumn{2}{c}{42} & \multicolumn{2}{c}{17} & \multicolumn{2}{c||}{13} & $^1$\ce{HCl} & 200 & 319 & 175 & \multicolumn{2}{c}{70} & \multicolumn{2}{c}{27} & 21 & $^1$\ce{ClF} & 350 & 415 & 217 & \multicolumn{2}{c}{98} & \multicolumn{2}{c}{29} & \multicolumn{2}{c}{23}\tabularnewline
$^1$\ce{CN-} & 40 & 229 & 169 & \multicolumn{2}{c}{42} & \multicolumn{2}{c}{17} & \multicolumn{2}{c||}{13} & $^2$\ce{CP} & 200 & 307 & 175 & \multicolumn{2}{c}{70} & \multicolumn{2}{c}{27} & 21 & $^1$\ce{SiS} & 300 & 475 & 259 & \multicolumn{2}{c}{98} & \multicolumn{2}{c}{31} & \multicolumn{2}{c}{23}\tabularnewline
$^1$\ce{N2} & 40 & 229 & 169 & \multicolumn{2}{c}{42} & \multicolumn{2}{c}{17} & \multicolumn{2}{c||}{11} & $^1$\ce{CP-} & 300 & 409 & 217 & \multicolumn{2}{c}{70} & \multicolumn{2}{c}{27} & 21 & $^1$\ce{P2} & 400 & 499 & 259 & \multicolumn{2}{c}{98} & \multicolumn{2}{c}{31} & \multicolumn{2}{c}{23}\tabularnewline
$^1$\ce{NO+} & 40 & 235 & 169 & \multicolumn{2}{c}{42} & \multicolumn{2}{c}{17} & \multicolumn{2}{c||}{13} & $^1$\ce{CS} & 200 & 307 & 175 & \multicolumn{2}{c}{70} & \multicolumn{2}{c}{27} & 21 & $^3$\ce{PS-} & 350 & 481 & 259 & \multicolumn{2}{c}{98} & \multicolumn{2}{c}{31} & \multicolumn{2}{c}{23}\tabularnewline
$^3$\ce{NO-}$^{a}$ & 200 & 319 & 175 & \multicolumn{2}{c}{70} & \multicolumn{2}{c}{19} & \multicolumn{2}{c||}{15} & $^2$\ce{SiN} & 200 & 307 & 175 & \multicolumn{2}{c}{70} & \multicolumn{2}{c}{27} & 21 & $^3$\ce{S2} & 350 & 481 & 259 & \multicolumn{2}{c}{98} & \multicolumn{2}{c}{31} & \multicolumn{2}{c}{23}\tabularnewline
$^1$\ce{CO} & 40 & 229 & 169 & \multicolumn{2}{c}{70} & \multicolumn{2}{c}{17} & \multicolumn{2}{c||}{13} & $^1$\ce{SiN-} & 200 & 307 & 175 & \multicolumn{2}{c}{70} & \multicolumn{2}{c}{27} & 21 & $^1$\ce{SCl-} & 400 & 487 & 259 & \multicolumn{2}{c}{98} & \multicolumn{2}{c}{33} & \multicolumn{2}{c}{25}\tabularnewline
$^3$\ce{O2} & 200 & 325 & 175 & \multicolumn{2}{c}{70} & \multicolumn{2}{c}{19} & \multicolumn{2}{c||}{13} & $^1$\ce{NP} & 200 & 313 & 175 & \multicolumn{2}{c}{70} & \multicolumn{2}{c}{27} & 21 & $^1$\ce{Cl2} & 400 & 493 & 259 & \multicolumn{2}{c}{98} & \multicolumn{2}{c}{37} & \multicolumn{2}{c}{29}\tabularnewline
$^1$\ce{CF+} & 40 & 223 & 169 & \multicolumn{2}{c}{70} & \multicolumn{2}{c}{19} & \multicolumn{2}{c||}{15} & $^3$\ce{SN-} & 350 & 415 & 217 & \multicolumn{2}{c}{98} & \multicolumn{2}{c}{29} & 21 & $^2$\ce{Cl2-} & 400 & 505 & 277 & \multicolumn{2}{c}{98} & \multicolumn{2}{c}{37} & \multicolumn{2}{c}{29}\tabularnewline
$^3$\ce{CF-} & 200 & 313 & 175 & \multicolumn{2}{c}{70} & \multicolumn{2}{c}{21} & \multicolumn{2}{c||}{15} & \multicolumn{1}{l}{} &  &  & \multicolumn{1}{l}{} & \multicolumn{2}{c}{} & \multicolumn{2}{c}{} & \multicolumn{1}{l}{} & \multicolumn{1}{l}{} &  &  & \multicolumn{1}{l}{} & \multicolumn{2}{c}{} & \multicolumn{2}{c}{} & \multicolumn{2}{c}{}\tabularnewline
\end{tabular}
\par\end{centering}
\caption{Comparison of the smallest numerical basis sets $(r_{\infty},n_{\mu},n_{\nu})$
yielding at least six correct decimals by rounding in \textsc{x2dhf}
calculations on main-group molecules, taken from \citeref{Jensen2005},
versus the $\epsilon=10^{-10}$ proxy basis sets reproduced by \textsc{HelFEM}
with $r_{\infty}=40$ with $n_{\mu}$ radial functions and $l_{\sigma}$
and $l_{\pi}$ partial waves, which reproduce the reference values
of \citeref{Jensen2005} beyond microhartree accuracy (present work).
$^{a}$Reference \citeref{Jensen2005} erroneously reports a singlet
state for \ce{NO-}.\label{tab:Basis-MG}}
\end{sidewaystable*}

\begin{sidewaystable*}
\begin{centering}
\begin{tabular}{l|lll|r@{\extracolsep{0pt}.}lr@{\extracolsep{0pt}.}lr@{\extracolsep{0pt}.}lr@{\extracolsep{0pt}.}l||l|lll|r@{\extracolsep{0pt}.}lr@{\extracolsep{0pt}.}lll||l|lll|r@{\extracolsep{0pt}.}lr@{\extracolsep{0pt}.}lr@{\extracolsep{0pt}.}lr@{\extracolsep{0pt}.}l}
 & \multicolumn{3}{c|}{\textsc{x2dhf}} & \multicolumn{8}{c||}{\textsc{HelFEM}} &  & \multicolumn{3}{c|}{\textsc{x2dhf}} & \multicolumn{6}{c||}{\textsc{HelFEM}} &  & \multicolumn{3}{c|}{\textsc{x2dhf}} & \multicolumn{8}{c}{\textsc{HelFEM}}\tabularnewline
molecule & $r_{\infty}$ & $n_{\mu}$ & $n_{\nu}$ & \multicolumn{2}{c}{$n_{\mu}$} & \multicolumn{2}{c}{$l_{\sigma}$} & \multicolumn{2}{c}{$l_{\pi}$} & \multicolumn{2}{c||}{$l_{\delta}$} & molecule & $r_{\infty}$ & $n_{\mu}$ & $n_{\nu}$ & \multicolumn{2}{c}{$n_{\mu}$} & \multicolumn{2}{c}{$l_{\sigma}$} & $l_{\pi}$ & $l_{\delta}$ & molecule & $r_{\infty}$ & $n_{\mu}$ & $n_{\nu}$ & \multicolumn{2}{c}{$n_{\mu}$} & \multicolumn{2}{c}{$l_{\sigma}$} & \multicolumn{2}{c}{$l_{\pi}$} & \multicolumn{2}{c}{$l_{\delta}$}\tabularnewline
\hline 
\hline 
$^1$\ce{ScCl} & 350 & 541 & 295 & \multicolumn{2}{c}{98} & \multicolumn{2}{c}{38} & \multicolumn{2}{c}{28} & \multicolumn{2}{c||}{24} & $^2$\ce{TiN} & 350 & 577 & 295 & \multicolumn{2}{c}{98} & \multicolumn{2}{c}{34} & 26 & 22 & $^1$\ce{NiC} & 350 & 571 & 295 & \multicolumn{2}{c}{126} & \multicolumn{2}{c}{38} & \multicolumn{2}{c}{28} & \multicolumn{2}{c}{24}\tabularnewline
$^1$\ce{ScF} & 250 & 409 & 229 & \multicolumn{2}{c}{98} & \multicolumn{2}{c}{34} & \multicolumn{2}{c}{26} & \multicolumn{2}{c||}{22} & $^1$\ce{TiN+} & 350 & 571 & 295 & \multicolumn{2}{c}{98} & \multicolumn{2}{c}{34} & 26 & 22 & $^1$\ce{NiSi} & 400 & 619 & 325 & \multicolumn{2}{c}{126} & \multicolumn{2}{c}{42} & \multicolumn{2}{c}{32} & \multicolumn{2}{c}{26}\tabularnewline
$^1$\ce{ScH} & 350 & 565 & 295 & \multicolumn{2}{c}{98} & \multicolumn{2}{c}{34} & \multicolumn{2}{c}{26} & \multicolumn{2}{c||}{22} & $^3$\ce{VO-} & 350 & 571 & 295 & \multicolumn{2}{c}{98} & \multicolumn{2}{c}{34} & 26 & 22 & $^1$\ce{CuH} & 350 & 583 & 295 & \multicolumn{2}{c}{98} & \multicolumn{2}{c}{36} & \multicolumn{2}{c}{28} & \multicolumn{2}{c}{24}\tabularnewline
$^2$\ce{ScN+} & 350 & 565 & 295 & \multicolumn{2}{c}{98} & \multicolumn{2}{c}{34} & \multicolumn{2}{c}{26} & \multicolumn{2}{c||}{22} & $^3$\ce{CrC} & 350 & 571 & 295 & \multicolumn{2}{c}{98} & \multicolumn{2}{c}{36} & 26 & 22 & $^1$\ce{CuF} & 400 & 637 & 325 & \multicolumn{2}{c}{126} & \multicolumn{2}{c}{40} & \multicolumn{2}{c}{30} & \multicolumn{2}{c}{26}\tabularnewline
$^1$\ce{ScN} & 250 & 415 & 229 & \multicolumn{2}{c}{98} & \multicolumn{2}{c}{34} & \multicolumn{2}{c}{26} & \multicolumn{2}{c||}{22} & $^1$\ce{CrMn+}$^{a}$ & 400 & 601 & 325 & \multicolumn{2}{c}{126} & \multicolumn{2}{c}{44} & 32 & 28 & $^1$\ce{CuCl} & 350 & 547 & 295 & \multicolumn{2}{c}{126} & \multicolumn{2}{c}{44} & \multicolumn{2}{c}{32} & \multicolumn{2}{c}{28}\tabularnewline
$^1$\ce{ScO+} & 350 & 571 & 295 & \multicolumn{2}{c}{98} & \multicolumn{2}{c}{34} & \multicolumn{2}{c}{24} & \multicolumn{2}{c||}{22} & $^3$\ce{MnC-}$^{a}$ & 350 & 571 & 295 & \multicolumn{2}{c}{98} & \multicolumn{2}{c}{36} & 26 & 22 & $^1$\ce{Cu2} & 425 & 685 & 361 & \multicolumn{2}{c}{126} & \multicolumn{2}{c}{46} & \multicolumn{2}{c}{34} & \multicolumn{2}{c}{28}\tabularnewline
$^2$\ce{ScO} & 250 & 415 & 229 & \multicolumn{2}{c}{98} & \multicolumn{2}{c}{34} & \multicolumn{2}{c}{26} & \multicolumn{2}{c||}{22} & $^3$\ce{FeC} & 350 & 571 & 295 & \multicolumn{2}{c}{98} & \multicolumn{2}{c}{38} & 28 & 24 & $^1$\ce{CuLi} & 400 & 607 & 325 & \multicolumn{2}{c}{126} & \multicolumn{2}{c}{46} & \multicolumn{2}{c}{34} & \multicolumn{2}{c}{28}\tabularnewline
$^1$\ce{ScS-} & 350 & 541 & 295 & \multicolumn{2}{c}{98} & \multicolumn{2}{c}{38} & \multicolumn{2}{c}{28} & \multicolumn{2}{c||}{24} & $^1$\ce{CoC-} & 350 & 577 & 295 & \multicolumn{2}{c}{98} & \multicolumn{2}{c}{36} & 28 & 24 & $^2$\ce{ZnH} & 350 & 571 & 295 & \multicolumn{2}{c}{126} & \multicolumn{2}{c}{40} & \multicolumn{2}{c}{30} & \multicolumn{2}{c}{24}\tabularnewline
$^2$\ce{ScS} & 350 & 547 & 295 & \multicolumn{2}{c}{98} & \multicolumn{2}{c}{38} & \multicolumn{2}{c}{28} & \multicolumn{2}{c||}{24} & $^3$\ce{CoO-} & 400 & 643 & 325 & \multicolumn{2}{c}{98} & \multicolumn{2}{c}{38} & 28 & 24 & $^2$\ce{ZnF} & 350 & 565 & 295 & \multicolumn{2}{c}{126} & \multicolumn{2}{c}{42} & \multicolumn{2}{c}{30} & \multicolumn{2}{c}{26}\tabularnewline
\end{tabular}
\par\end{centering}
\caption{Comparison of the smallest numerical basis sets $(r_{\infty},n_{\mu},n_{\nu})$
yielding the energy converged to microhartrees, taken from \citeref{Williams2008},
versus the $\epsilon=10^{-10}$ proxy basis sets reproduced by \textsc{HelFEM}
with $r_{\infty}=40$ with $n_{\mu}$ radial functions and $l_{\sigma}$
and $l_{\pi}$ partial waves, which reproduce the reference values
beyond microhartree accuracy (present work). $^{a}$The reference
values reported in \citeref{Williams2008} are incorrect for \ce{CrMn+}
and \ce{MnC-}, see main text.\label{tab:Basis-TM}}
\end{sidewaystable*}

Having shown that the proxy basis sets are capable of reproducing
energies at the ROHF limit, for reference, updated ROHF reference
values, computed in the $r_{\infty}=60a_{0}$, $\epsilon=10^{-10}$
basis set for the 70 diatomic molecules of \citerefs{Jensen2005}
and \citenum{Williams2008} are shown in \tabref{Updated-restricted-open-shell}.
In most cases, the change is but a different round-off of the last
decimal; however, the changes to the ROHF limit energy of $^{3}$\ce{SiH-}
and $^{2}$\ce{Cl2-} as well as the aforementioned $^{1}$\ce{CrMn+}
and $^{3}$\ce{MnC-} are more noticeable. In addition, we have repeated
the calculations with unrestricted HF for the molecules for which
non-singlet states were specified; these results are shown in \tabref{Unrestricted-Hartree=002013Fock-result}.
The energy lowerings from ROHF range from 0.44 m$E_{h}$ for ScO to
241.6 m$E_{h}$ for \ce{FeC}; these results were likewise obtained
with the $r_{\infty}=60a_{0}$, $\epsilon=10^{-10}$ basis set.

\begin{table*}
\begin{centering}
\begin{tabular}{lrllrllr}
molecule & energy &  & molecule & energy &  & molecule & energy\tabularnewline
\hline 
\hline 
$^{3}$\ce{NH} & -54.978423\textbf{8} &  & $^{2}$\ce{CP} & -378.474608\textbf{3} &  & $^{3}$\ce{PS-} & -738.339707\textbf{0}\tabularnewline
$^{1}$\ce{OH-} & -75.418803\textbf{0} &  & $^{1}$\ce{CP-} & -378.561588\textbf{6} &  & $^{3}$\ce{S2} & -795.091559\textbf{1}\tabularnewline
$^{1}$\ce{CN-} & -92.348950\textbf{5} &  & $^{1}$\ce{CS} & -435.362420\textbf{1} &  & $^{1}$\ce{SCl-} & -857.104418\textbf{4}\tabularnewline
$^{1}$\ce{NO+} & -128.978051\textbf{4} &  & $^{2}$\ce{SiN} & -343.297026\textbf{8} &  & $^{1}$\ce{Cl2} & -919.008934\textbf{8}\tabularnewline
$^{3}$\ce{NO-} & -129.280174\textbf{6} &  & $^{1}$\ce{SiN-} & -343.362365\textbf{5} &  & $^{2}$\ce{Cl2-} & -919.07956\textbf{46}\tabularnewline
$^{3}$\ce{O2} & -149.668757\textbf{3} &  & $^{1}$\ce{NP} & -395.188395\textbf{3} &  & $^{1}$\ce{ScCl} & -1219.33578\textbf{54}\tabularnewline
$^{3}$\ce{CF-} & -137.224456\textbf{1} &  & $^{3}$\ce{SN-} & -451.987649\textbf{2} &  & $^{1}$\ce{ScO+} & -834.44152\textbf{32}\tabularnewline
$^{3}$\ce{NF} & -153.842421\textbf{1} &  & $^{3}$\ce{NCl} & -513.907013\textbf{4} &  & $^{1}$\ce{CrMn+} & -2192.\textbf{3703627}\tabularnewline
$^{1}$\ce{OF-} & -174.236341\textbf{7} &  & $^{1}$\ce{SiO} & -363.855341\textbf{6} &  & $^{3}$\ce{MnC-} & -1187.32\textbf{40934}\tabularnewline
$^{1}$\ce{F2} & -198.773444\textbf{9} &  & $^{1}$\ce{ClF} & -558.917626\textbf{4} &  & $^{3}$\ce{CoO-} & -1456.14375\textbf{65}\tabularnewline
$^{3}$\ce{SiH-} & -289.4646\textbf{299} &  & $^{1}$\ce{SiS} & -686.516284\textbf{0} &  & $^{1}$\ce{Cu2} & -3277.94160\textbf{67}\tabularnewline
$^{1}$\ce{SH-} & -398.149790\textbf{8} &  &  &  &  &  & \tabularnewline
\end{tabular}
\par\end{centering}
\caption{ROHF limit values updated from \tabref{List-MG, List-TM} using the
corresponding $r_{\infty}=60a_{0}$, $\epsilon=10^{-10}$ basis set,
with changed digits shown in bold.\label{tab:Updated-restricted-open-shell}}
\end{table*}

\begin{table}
\begin{centering}
\begin{tabular}{lr@{\extracolsep{0pt}.}llr@{\extracolsep{0pt}.}llr@{\extracolsep{0pt}.}l}
molecule & \multicolumn{2}{c}{energy} & molecule & \multicolumn{2}{c}{energy} & molecule & \multicolumn{2}{c}{energy}\tabularnewline
\hline 
\hline 
\ce{CH-} & -38&2994602 & \ce{SiN} & -343&3130427 & \ce{ScO} & -834&6749525\tabularnewline
\ce{NH} & -54&9863336 & \ce{SN-} & -452&0046200 & \ce{ScS} & -1157&3394254\tabularnewline
\ce{CN} & -92&2425169 & \ce{NCl} & -513&9187494 & \ce{TiN} & -902&7833285\tabularnewline
\ce{NO-} & -129&2959898 & \ce{PO-} & -415&6654029 & \ce{VO-} & -1017&7882900\tabularnewline
\ce{O2} & -149&6922860 & \ce{SO} & -472&4170546 & \ce{CrC} & -1081&0323244\tabularnewline
\ce{CF-} & -137&2322589 & \ce{PF} & -440&2409912 & \ce{MnC-} & -1187&5638036\tabularnewline
\ce{NF} & -153&8527981 & \ce{PS-} & -738&3489562 & \ce{FeC} & -1300&1679109\tabularnewline
\ce{F2-} & -198&8783311 & \ce{S2} & -795&1075610 & \ce{CoO-} & -1456&2267835\tabularnewline
\ce{SiH-} & -289&4704308 & \ce{Cl2-} & -919&0880177 & \ce{ZnH} & -1778&3801627\tabularnewline
\ce{CP} & -378&4754390 & \ce{ScN+} & -813&9079564 & \ce{ZnF} & -1877&3457934\tabularnewline
\end{tabular}
\par\end{centering}
\caption{Unrestricted HF energies for the main group and transition metal systems
obtained with the corresponding $r_{\infty}=60a_{0}$, $\epsilon=10^{-10}$
basis set, using the geometries and spin states given in \tabref{List-MG, List-TM}.\label{tab:Unrestricted-Hartree=002013Fock-result}}
\end{table}

\subsection{Electric properties of \ce{BH} and \ce{N2}\label{subsec:Electric-properties-of}}

Next, to demonstrate further capabilities of the program, we run finite
field HF calculations on the \ce{BH} ($R=2.3289a_{0}$) and \ce{N2}
(\textbf{$R=2.068a_{0}$}) molecules and compare the results with
literature values computed at the basis set limit from \citeref{Kobus2015},
complemented with unpublished data from the same work.\citep{Kobus-communication}
Five radial elements are used with $r_{\infty}=40a_{0}$ with the
angular basis $l_{\sigma}=20$, $l_{\pi}=15$ for both molecules;
this yields the values in \tabref{Electric-properties}. For comparison,
the basis with $\epsilon=10^{-10}$ would have three radial elements
and $l_{\sigma}=17$, $l_{\pi}=11$ for \ce{N2} and $l_{\sigma}=15$,
$l_{\pi}=11$ for \ce{BH} at the used geometries.

For BH, the energies in \tabref{Electric-properties} match to nanohartree-level
accuracy. The dipole and quadrupole moments only disagree in the fifth
and sixth decimals, respectively, again indicating an excellent level
of agreement. What makes this remarkable is that \citeref{Kobus2015}
employed 349 points in $\nu$ and 643 points in $\mu$ with $r_{\infty}=200a_{0}$,
whereas the calculations in the present work employ but 21 $\sigma$
waves and $15$ $\pi$ waves, and 70 shape functions in $\mu$ with
$r_{\infty}=40a_{0}$. That is, we obtain excellent accuracy despite
having used\emph{ over two orders of magnitude} fewer parameters (a
factor of over 150) for the wave function, in line with the results
that were obtained above in \subsecref{HF-limit-energies} for the
field-free case.

For \ce{N2}, the energies again agree to nanohartree-level accuracy,
whereas differences in the dipole and quadrupole moments are now seen
already at the fourth and fifth decimals, respectively. \Citeref{Kobus2015}
employed an even larger grid for \ce{N2} than for BH: 841 points
in $\mu$ and 445 points in $\nu$, whereas the present calculations
only use the same small number of parameters as for \ce{BH}. Again,
a reduction of over two orders of magnitude is achieved, underlining
the power of the present approach.

Employing the data in \tabref{Electric-properties}, we obtain the
polarizabilities 22.561246 and 22.560787 for BH and 14.950727 and
14.949617 for \ce{N2}, employing the two-point 
\begin{equation}
f'(x)\approx\frac{f(x+h)-f(x-h)}{2h}\label{eq:twopoint}
\end{equation}
and four-point 
\begin{equation}
f'(x)\approx\frac{-f(x+2h)+8f(x+h)-8f(x-h)+f(x-2h)}{12h}\label{eq:fourpoint}
\end{equation}
stencils, respectively. These results compare favorably with the literature
values\citep{Kobus2015} 22.560640 for BH and 14.9512154 for \ce{N2}:
the discrepancy for the four-point value for BH is at the sixth significant
number -- well within the estimated numerical error bounds -- whereas
for \ce{N2} the discrepancy is already seen at the fifth significant
number, still yielding good agreement.

\begin{sidewaystable*}
\begin{centering}
\begin{tabular}{r@{\extracolsep{0pt}.}lr@{\extracolsep{0pt}.}lr@{\extracolsep{0pt}.}lr@{\extracolsep{0pt}.}lr@{\extracolsep{0pt}.}lr@{\extracolsep{0pt}.}lr@{\extracolsep{0pt}.}lr@{\extracolsep{0pt}.}l}
\multicolumn{16}{c}{BH, $R=2.3289a_{0}$}\tabularnewline
\multicolumn{2}{c}{} & \multicolumn{6}{c}{{\footnotesize{}Present work}} & \multicolumn{2}{c}{} & \multicolumn{6}{c}{{\footnotesize{}Literature value}}\tabularnewline
\multicolumn{2}{c}{{\footnotesize{}$E_{z}$}} & \multicolumn{2}{c}{{\footnotesize{}Energy$^{a}$}} & \multicolumn{2}{c}{{\footnotesize{}Dipole}} & \multicolumn{2}{c}{{\footnotesize{}Quadrupole}} & \multicolumn{2}{c}{} & \multicolumn{2}{c}{{\footnotesize{}Energy$^{b}$}} & \multicolumn{2}{c}{{\footnotesize{}Dipole$^{b}$}} & \multicolumn{2}{c}{{\footnotesize{}Quadrupole$^{c}$}}\tabularnewline
\hline 
\hline 
\multicolumn{2}{c}{$-8\times10^{-4}$} & -25&132024018 & 6&669105 (-1) & -3&968475 (0) & \multicolumn{2}{c}{} & -25&132024018 & 6&669096 (-1) & -3&968472 (0)\tabularnewline
\multicolumn{2}{c}{$-4\times10^{-4}$} & -25&131829783 & 6&759386 (-1) & -3&985054 (0) & \multicolumn{2}{c}{} & -25&131829783 & 6&759377 (-1) & -3&985051 (0)\tabularnewline
\multicolumn{2}{c}{0} & -25&131639159 & 6&849639 (-1) & -4&001849 (0) & \multicolumn{2}{c}{} & -25&131639159 & 6&849630 (-1) & -4&001846 (0)\tabularnewline
\multicolumn{2}{c}{$4\times10^{-4}$} & -25&131452145 & 6&939876 (-1) & -4&018863 (0) & \multicolumn{2}{c}{} & -25&131452145 & 6&939866 (-1) & -4&018861 (0)\tabularnewline
\multicolumn{2}{c}{$8\times10^{-4}$} & -25&131268739 & 7&030107 (-1) & -4&036098 (0) & \multicolumn{2}{c}{} & -25&131268740 & 7&030097 (-1) & -4&036096 (0)\tabularnewline
\multicolumn{2}{c}{} & \multicolumn{2}{c}{} & \multicolumn{2}{c}{} & \multicolumn{2}{c}{} & \multicolumn{2}{c}{} & \multicolumn{2}{c}{} & \multicolumn{2}{c}{} & \multicolumn{2}{c}{}\tabularnewline
\multicolumn{16}{c}{\ce{N2}, $R=2.068a_{0}$}\tabularnewline
\multicolumn{2}{c}{} & \multicolumn{6}{c}{{\footnotesize{}Present work}} & \multicolumn{2}{c}{} & \multicolumn{6}{c}{{\footnotesize{}Literature value}}\tabularnewline
\multicolumn{2}{c}{{\footnotesize{}$E_{z}$$^{d}$}} & \multicolumn{2}{c}{{\footnotesize{}Energy}} & \multicolumn{2}{c}{{\footnotesize{}Dipole}} & \multicolumn{2}{c}{{\footnotesize{}Quadrupole}} & \multicolumn{2}{c}{} & \multicolumn{2}{c}{{\footnotesize{}Energy$^{b}$}} & \multicolumn{2}{c}{{\footnotesize{}Dipole$^{b}$}} & \multicolumn{2}{c}{{\footnotesize{}Quadrupole$^{b}$}}\tabularnewline
\hline 
\hline 
\multicolumn{2}{c}{$-1.6\times10^{-3}$} & -108&993844772 & -2&392544 (-2) & -9&401255 (-1) & \multicolumn{2}{c}{} & -108&993844772 & -2&392249 (-2) & -9&401254 (-1)\tabularnewline
\multicolumn{2}{c}{$-0.8\times10^{-3}$} & -108&993830419 & -1&196071 (-2) & -9&399594 (-1) & \multicolumn{2}{c}{} & -108&993830419 & -1&196104 (-2) & -9&399567 (-1)\tabularnewline
\multicolumn{2}{c}{0} & -108&993825634 & -9&782586 (-13) & -9&399016 (-1) & \multicolumn{2}{c}{} & -108&993825635 & -3&463896 (-14) & -9&399005 (-1)\tabularnewline
\multicolumn{2}{c}{$0.8\times10^{-3}$} & -108&993830419 & 1&196071 (-2) & -9&399594 (-1) & \multicolumn{2}{c}{} & -108&993830419 & 1&196104 (-2) & -9&399567 (-1)\tabularnewline
\multicolumn{2}{c}{$1.6\times10^{-3}$} & -108&993844772 & 2&392544 (-2) & -9&401255 (-1) & \multicolumn{2}{c}{} & -108&993844772 & 2&392249 (-2) & -9&401254 (-1)\tabularnewline
\end{tabular}
\par\end{centering}
\centering{}\caption{Electric properties of \ce{BH} and \ce{N2} in a finite field, compared
to literature values. The values in the parentheses indicate magnitude,
$A(n)=A\times10^{n}$.\protect \\
$^{a}$The energy for \textsc{HelFEM} is offset to match that of \citeref{Kobus2015},
due to a difference in the definition of the zero-point of the nuclear
dipole -- electric field interaction.\protect \\
$^{b}$Literature values are from \citeref{Kobus2015} with a truncated
number of decimals.\protect \\
$^{c}$The values for the quadrupole moment of BH with respect to
the center of the molecule for were obtained from \citeref{Kobus-communication}.\protect \\
$^{d}$Note that the values for the field reported in Table V of \citeref{Kobus2015}
correspond in fact to the $F_{z}$ values given above instead of $\pm0.6\times10^{-3}$
and $\pm1.2\times10^{-3}$.\citep{Kobus-communication}\label{tab:Electric-properties}}
\end{sidewaystable*}

\subsection{Atomization energy of \ce{N2}}

As a final demonstration, we study the convergence of the atomization
energy
\begin{equation}
\Delta E=\sum_{\text{atoms }i}E_{i}^{\text{atom}}-E^{\text{molecule}}>0\label{eq:atomization}
\end{equation}
of \ce{N2} at the geometry given in \tabref{Electric-properties}
with HF, and the LDA,\citep{Bloch1929,Dirac1930,Perdew1992a} PBE,\citep{Perdew1996,Perdew1997}
PBE0,\citep{Adamo1999,Ernzerhof1999} BP86,\citep{Becke1988a,Perdew1986}
BLYP,\citep{Becke1988a,Lee1988} B3LYP,\citep{Becke1993,Stephens1994}
revTPSS,\citep{Perdew2009,Perdew2011} revTPSSh,\citep{Perdew2009,Perdew2011,Csonka2010}
MS2,\citep{Sun2013a,Perdew2009} and MS2h\citep{Sun2013a,Perdew2009}
functionals. Although atomic energies can be computed most efficiently
with the atomic program presented in part I,\cite{Lehtola2019a} the
atomization energy can be extracted more accurately by running the
atomic calculations in the same basis set as the diatomic molecule,
achieved by setting the other nuclear charge to 0, in analogy to the
Boys--Bernardi counterpoise method of LCAO calculations.\citep{Boys1970}
Such a procedure results in significant error cancellation: the largest
error in the total energy for both the individual atoms and the molecule
arises from an incomplete description in $\nu$ of the core region,
which requires many partial waves to converge fully. By computing
the atomic energies in the same basis set, the errors arising from
the core region cancel out almost perfectly.

The results with this method are shown in \tabref{atE-partial}, highlighting
excellent, monotonic convergence for all methods. Note that unlike
\figref{conv-mg,conv-tm}, \tabref{atE-partial} does not contain
values for arbitrary values of $\epsilon$, as the estimated error
in the proxy energy may decrease by several orders of magnitude per
step: for example, the $\epsilon=10^{-3}$ basis is the same as the
$\epsilon=10^{-4}$ basis. The atomization energy evaluated with the
fully numerical approach appears to be converged to 0.1 meV accuracy
with all methods except MS2 and MS2h, for which convergence is slightly
poorer; this can likely be attributed to the MS2 exchange functional
being numerically less well-conditioned than the other functionals
in the present study.

For comparison, \tabref{atE-gauss} shows the corresponding calculations
performed with \textsc{Erkale}\citep{erkale,Lehtola2012} in the pcseg-$n$
and aug-pcseg-$n$ basis sets.\citep{Jensen2001,Jensen2002b,Jensen2014}
A (350,974) DFT quadrature grid and the Boys--Bernardi counterpoise
correction\citep{Boys1970} were employed to ensure benchmark quality
results. The largest differences between the \textsc{HelFEM} and \textsc{Erkale}
results with the best basis sets, $\epsilon=10^{-10}$ and aug-pcseg-4,
respectively, are seen for the MS2 (43.7 meV) and MS2h (40.2 meV)
functionals, again likely caused by the numerical properties of the
MS2 exchange functional. The revTPSS and revTPSSh values disagree
by 12.6 meV and 10.6 meV, respectively. The disagreements for the
other functionals are in the range of 2--6 meV. Although the number
of basis functions in the Gaussian basis calculations is much smaller
than in the partial wave approach, convergence with the Gaussian basis
sets is not always monotonic, unlike what was observed in \tabref{atE-partial}
for the partial wave method.

\begin{sidewaystable*}
\begin{tabular}{llr@{\extracolsep{0pt}.}lr@{\extracolsep{0pt}.}lr@{\extracolsep{0pt}.}lr@{\extracolsep{0pt}.}lr@{\extracolsep{0pt}.}lr@{\extracolsep{0pt}.}lr@{\extracolsep{0pt}.}lr@{\extracolsep{0pt}.}lr@{\extracolsep{0pt}.}lr@{\extracolsep{0pt}.}lr@{\extracolsep{0pt}.}l}
basis & $n_{\text{bf}}$ & \multicolumn{2}{c}{HF} & \multicolumn{2}{c}{LDA} & \multicolumn{2}{c}{PBE} & \multicolumn{2}{c}{PBE0} & \multicolumn{2}{c}{BP86} & \multicolumn{2}{c}{BLYP} & \multicolumn{2}{c}{B3LYP} & \multicolumn{2}{c}{revTPSS} & \multicolumn{2}{c}{revTPSSh} & \multicolumn{2}{c}{MS2} & \multicolumn{2}{c}{MS2h}\tabularnewline
$10^{-0}$ & 190 & 9&5597 & 16&3779 & 15&0682 & 14&2955 & 15&1409 & 15&0568 & 14&6154 & 14&4547 & 14&1898 & 14&4645 & 14&2433\tabularnewline
$10^{-1}$ & 242 & 5&2812 & 11&9074 & 10&8646 & 10&0738 & 10&8872 & 10&7276 & 10&2628 & 10&2409 & 9&9694 & 10&4266 & 10&1814\tabularnewline
$10^{-2}$ & 830 & 5&1822 & 11&8001 & 10&7499 & 9&9632 & 10&7757 & 10&6138 & 10&1523 & 10&1173 & 9&8482 & 10&3602 & 10&1108\tabularnewline
$10^{-4}$ & 1078 & 5&0307 & 11&6257 & 10&5800 & 9&7979 & 10&6053 & 10&4405 & 9&9826 & 9&9589 & 9&6903 & 10&1846 & 9&9375\tabularnewline
$10^{-7}$ & 1326 & 5&0268 & 11&6216 & 10&5751 & 9&7932 & 10&6004 & 10&4357 & 9&9780 & 9&9533 & 9&6849 & 10&1760 & 9&9297\tabularnewline
$10^{-9}$ & 1574 & 5&0267 & 11&6215 & 10&5751 & 9&7932 & 10&6003 & 10&4356 & 9&9779 & 9&9532 & 9&6849 & 10&1759 & 9&9297\tabularnewline
$10^{-10}$ & 1658 & 5&0267 & 11&6215 & 10&5751 & 9&7932 & 10&6003 & 10&4356 & 9&9779 & 9&9531 & 9&6848 & 10&1752 & 9&9291\tabularnewline
\end{tabular}

\caption{Atomization energy of \ce{N2} in eV with HF and the LDA, PBE, PBE0,
BP86, BLYP, B3LYP, revTPSS, revTPSSh, MS2, and MS2h functionals, employing
the partial wave expansion with the adaptive grid method.\label{tab:atE-partial}}
\end{sidewaystable*}
\begin{sidewaystable*}
\begin{tabular}{llr@{\extracolsep{0pt}.}lr@{\extracolsep{0pt}.}lr@{\extracolsep{0pt}.}lr@{\extracolsep{0pt}.}lr@{\extracolsep{0pt}.}lr@{\extracolsep{0pt}.}lr@{\extracolsep{0pt}.}lr@{\extracolsep{0pt}.}lr@{\extracolsep{0pt}.}lr@{\extracolsep{0pt}.}lr@{\extracolsep{0pt}.}l}
basis & $n_{\text{bf}}$ & \multicolumn{2}{c}{HF} & \multicolumn{2}{c}{LDA} & \multicolumn{2}{c}{PBE} & \multicolumn{2}{c}{PBE0} & \multicolumn{2}{c}{BP86} & \multicolumn{2}{c}{BLYP} & \multicolumn{2}{c}{B3LYP} & \multicolumn{2}{c}{revTPSS} & \multicolumn{2}{c}{revTPSSh} & \multicolumn{2}{c}{MS2} & \multicolumn{2}{c}{MS2h}\tabularnewline
\hline 
\hline 
pcseg-0 & 18 & 2&4926 & 9&9488 & 8&9278 & 7&8865 & 8&9383 & 8&8661 & 8&1961 & 8&1253 & 7&7757 & 8&2910 & 7&9759\tabularnewline
pcseg-1 & 28 & 4&7316 & 11&2740 & 10&2694 & 9&4830 & 10&2788 & 10&0954 & 9&6412 & 9&6875 & 9&4147 & 9&8871 & 9&6381\tabularnewline
pcseg-2 & 60 & 4&9930 & 11&5768 & 10&5323 & 9&7514 & 10&5618 & 10&3910 & 9&9342 & 9&9190 & 9&6509 & 10&1092 & 9&8655\tabularnewline
pcseg-3 & 120 & 5&0343 & 11&6096 & 10&5572 & 9&7828 & 10&5876 & 10&4181 & 9&9670 & 9&9400 & 9&6744 & 10&1243 & 9&8831\tabularnewline
pcseg-4 & 202 & 5&0303 & 11&6179 & 10&5673 & 9&7887 & 10&5944 & 10&4289 & 9&9743 & 9&9387 & 9&6727 & 10&1303 & 9&8878\tabularnewline
aug-pcseg-0 & 26 & 2&3501 & 9&6348 & 8&5963 & 7&6248 & 8&6306 & 8&5292 & 7&9169 & 7&8346 & 7&5088 & 8&0288 & 7&7309\tabularnewline
aug-pcseg-1 & 46 & 4&7695 & 11&3205 & 10&2978 & 9&5209 & 10&3128 & 10&1309 & 9&6840 & 9&7068 & 9&4388 & 9&9076 & 9&6624\tabularnewline
aug-pcseg-2 & 92 & 5&0070 & 11&5865 & 10&5408 & 9&7616 & 10&5688 & 10&3990 & 9&9437 & 9&9264 & 9&6592 & 10&1171 & 9&8742\tabularnewline
aug-pcseg-3 & 170 & 5&0364 & 11&6177 & 10&5681 & 9&7898 & 10&5949 & 10&4290 & 9&9745 & 9&9475 & 9&6809 & 10&1316 & 9&8895\tabularnewline
aug-pcseg-4 & 274 & 5&0306 & 11&6194 & 10&5690 & 9&7897 & 10&5959 & 10&4308 & 9&9754 & 9&9405 & 9&6742 & 10&1315 & 9&8889\tabularnewline
\end{tabular}

\caption{Atomization energy of \ce{N2} in eV with HF and the LDA, PBE, PBE0,
BP86, BLYP, B3LYP, revTPSS, revTPSSh, MS2, and MS2h, employing (aug-)pcseg-$n$
basis sets.\label{tab:atE-gauss}}
\end{sidewaystable*}

\section{Summary}

We have presented a new finite element implementation of the partial
wave approach for diatomic molecules in the \textsc{HelFEM} program\citep{HelFEM}
for electronic structure calculations with Hartree--Fock (HF) or
density functional theory (DFT). \textsc{HelFEM} supports hundreds
of functionals within the local spin density approximation (LDA),
generalized gradient approximation (GGA), as well as meta-GGA approximation
-- including hybrid functionals -- via an interface to the \textsc{Libxc}
library.\citep{Lehtola2018} The orbitals can be fully spin-restricted,
spin-restricted open-shell, or fully spin-unrestricted in calculations.

We have proposed a novel way to cost-efficiently choose the fully
numerical basis set for calculations on diatomic molecules by optimizing
the completeness of the basis set to reproduce the lowest eigenstates
of the core Hamiltonian. By applying the procedure to calculations
on 70 diatomic molecules with published restricted open-shell HF limit
energies from the literature, we showed that the approach is able
to easily and controllably reproduce energies at a sub-microhartree
level accuracy, requiring a significantly smaller number of parameters
for the wave function than what was originally needed to generate
the literature values. Further applications of the program to the
electric properties of \ce{BH} and \ce{N2} under finite field also
showed excellent agreement with previously published values, even
though over two orders of magnitude fewer parameters were used for
the wave function in the present work. The application of the program
to the atomization energy of \ce{N2} with HF and local spin density
(LDA), generalized gradient approximation (GGA), and meta-GGA functionals
and comparison to Gaussian basis set calculations further underlined
the robustness of the present approach. The extension of the present
work to finite magnetic fields has been discussed in \citeref{Lehtola2019d}.

\section{Discussion}

Although many systems are already tractable with the present version
of \textsc{HelFEM}, it is evident that as a novel program, many further
optimizations are possible. This is especially clear from McCullough's
paper from over 30 years ago that reported a spin-restricted single-reference
calculation on the $^{2}\Pi$ state of \ce{KO} at $4.40a_{0}$ bond
length with $l_{\max}=29$, with the converged final energy -674.014150
$E_{h}$;\citep{McCullough1986} we have repeated the calculation
with the adaptive basis at an estimated $10^{-10}$ accuracy, yielding
$l_{\sigma}=37$ and $l_{\pi}=27$ with seven 15-node Lobatto elements,
yielding the final energy -674.014903429 $E_{h}$. As such large calculations
were possible already in the mid-1980s, the feasible system size limit
with present-day computers and algorithms should be much larger. The
venerable \textsc{x2dhf} program might also yield insights into possible
further optimizations.

At present, alike the atomic program presented in part I of the series,\cite{Lehtola2019a}
all matrices in the diatomic program are stored naïvely as dense matrices
with the rank $N_{\text{ang}}\times N_{\text{rad}}$, where $N_{\text{ang}}$
and $N_{\text{rad}}$ are the number of angular and radial basis functions,
as this is easier to implement and develop upon than a more specialized
storage scheme. However, as was stated by \eqref{diatom}, the orbitals
block by the $m$ quantum number, and so the orbital gradient is also
diagonal in $m$, unless symmetries are broken. This means that the
self-consistent field (SCF) problem could in principle be solved using
\emph{e.g.} DIIS by only building the $(m,m)$ blocks of the (Kohn--Sham--)Fock
matrix, which would mean a savings of a factor of roughly $2m_{\max}+1$
in the size of the matrix.

However, evaluations of the total energy require also the off-diagonal
$(m,m')$ blocks. As the DIIS method only works when the orbitals
are sufficiently close to convergence, more robust methods are required
for the initial steps of the SCF procedure, before switching over
to the DIIS algorithm. But, as the more robust algorithms such as
the presently used ADIIS\citep{Hu2010} algorithm typically require
evaluations of the total energy, the potential savings of not computing
the $(m,m')$ blocks for DIIS would be small: the bulk of SCF iterations
are typically spent on getting the orbitals close to convergence,
after which DIIS converges within a few iterations. The $m$ factor
is also quite small even for $\varphi$ orbitals, for which the saving
would be only a factor of 7.

\section*{Funding information}

This work has been supported by the Academy of Finland through project
number 311149.

\section*{Acknowledgments}

I thank Barry Schneider for invaluable help with Legendre functions,
and Dage Sundholm and Jacek Kobus for discussions. I also thank Dage
Sundholm, Barry Schneider, Jacek Kobus, and Pekka Pyykkö for comments
on the manuscript, as well as Frank Jensen and Jacek Kobus for help
in reproducing literature results. Computational resources provided
by CSC -- It Center for Science Ltd (Espoo, Finland) and the Finnish
Grid and Cloud Infrastructure (persistent identifier urn:nbn:fi:research-infras-2016072533)
are gratefully acknowledged.

\bibliographystyle{apsrev}
\bibliography{citations}

\end{document}